\documentclass[reprint,
aps,prd,amsmath,amssymb,nofootinbib,twocolumn]{revtex4-2}

\usepackage{graphicx}
\usepackage{dcolumn}
\usepackage{xcolor}
\usepackage{hyperref}
\usepackage{bm}
\usepackage{braket}
\usepackage{multirow}
\usepackage{booktabs}
\usepackage{array}
\usepackage{tabularx}
\usepackage{makecell}
\usepackage{tensor}

\hypersetup{colorlinks=true, linkcolor=blue, citecolor=blue, urlcolor=blue, filecolor=blue}

\begin{document}

\title{On Pulsar Timing Detection of Ultralight Vector Dark Matter}

\author{Jeff A. Dror}
\email{jeffdror@ufl.edu}
\affiliation{Institute for Fundamental Theory, Physics Department,\\ University of Florida, Gainesville, FL 32611, USA}

\author{Qiushi Wei}
\email{q.wei@ufl.edu}
\affiliation{Institute for Fundamental Theory, Physics Department,\\ University of Florida, Gainesville, FL 32611, USA}

\begin{abstract}

Ultralight vector dark matter induces metric fluctuations that generate timing residuals in the arrival times of pulsar emissions through two distinct modes: a fast mode, sourced by coherent field oscillations, and a slow mode, arising from interference patterns. These modes enable the detection of vector dark matter with masses $m \sim 10^{-24} - 10^{-22}\ \mathrm{eV}$ and $m \sim 10^{-18} - 10^{-16}\ \mathrm{eV}$, respectively, using pulsar timing arrays. While previous studies have explored the fast mode, they neglect the full statistical treatment of the vector field and a precise treatment of its polarization structure. In this work, we investigate the timing residuals from both modes, fully accounting for the statistical properties of ultralight vector dark matter, assuming equipartition among its three polarization states. The two-point correlation functions of timing residuals that we derive serve as direct tools for identifying vector dark matter signatures as a stochastic background in pulsar timing data.

\end{abstract}

\maketitle

\section{\label{sec1}Introduction}
If dark matter has a mass well below 1~eV, it behaves as a classical field within our galaxy due to its large state occupation number. 
Such wave-like dark matter induces oscillatory weak-field perturbations in the spacetime metric -- analogous to gravitational waves -- but with perturbations that extend beyond purely propagating tensor modes. 
The characteristics of these perturbations are governed by the properties of dark matter, most notably its mass, spin, and velocity distribution. 
These metric perturbations can be probed using techniques similar to those employed in gravitational wave detection, offering a promising avenue for {\em gravitational direct detection} of dark matter.

The metric perturbations generated by ultralight dark matter exhibit two characteristic modes, each with distinct observational signatures. 

The first, known as the {\em fast mode}, has an angular frequency of $\omega \simeq 2m$, where $m$ is the dark matter mass.  This mode arises from the coherent oscillation of the dark matter field at its Compton frequency and was first identified by Khmelnitsky and Rubakov~\cite{Rubakov:2014} in the context of scalar dark matter detection via pulsar timing. 
Efforts to detect the fast mode of ultralight dark matter have been pursued using laser interferometers~\cite{Aoki:2016_interferometer,Yong:2024_interferometer}, pulsar timing~\cite{Porayko:2014_PTA,Graham:2015_PTA,Aoki:2016_PTA,DeMartino:2017_PTA,Kato:2019_PTA,Nomura:2019_PTA,Unal:2022_PTA,Xia:2023_PTA,Luu:2023_PTA,Hwang:2023_PTA,Nomura:2023_PTA,Dror:2025_PTA}, astrometry~\cite{Chen:2022_Astrometry,Dror:2024_astrometry,Kim:2024_astrometry,An:2024_astrometry}, and through its influence on gravitational wave observations of other astrophysical sources~\cite{Blas:2016_onGW,Wang:2023_onGW,Delgado:2023_onGW,Brax:2024_onGW,Blas:2024_onGW}. 

The second, known as the {\em slow mode}, has angular frequencies $\omega \lesssim mv_0^2$, where $v_0 \sim 10^{-3}$ is the local velocity dispersion of dark matter.~\footnote{Specifically, we assume an isotropic velocity distribution in the halo, and $v_0$ is defined as the standard deviation of one component of the velocity. It also represents the typical velocity of the local dark matter. This definition is general; we will specify it further in the Maxwell-Boltzmann distribution later.}
This mode arises from the interference pattern of the stochastic dark matter field and is often referred to in previous literature as de Broglie scale ``granules'' in a halo~\cite{Hui:2021,Schive:2014}.
Although it has been discussed regarding its modulation on the fast mode signals~\cite{DeMartino:2017_PTA, Luu:2023_PTA}, its own detectability has only recently been explored for scalar dark matter using interferometers~\cite{Kim:2023_interferometer}, pulsar timing arrays~\cite{Kim:2023_PTA,Eberhardt:2024_PTA}, and astrometry~\cite{Kim:2024_astrometry}. 

The amplitudes of both fast and slow mode perturbations scale with the ratio of the time-averaged energy density to the dark matter mass squared. 
Thus, for fixed energy density, lighter dark matter induces larger perturbations. Pulsar timing arrays, which probe metric oscillations at nanohertz frequencies,~\footnote{This is a modest oversimplification since pulsar timing arrays can probe frequencies well outside this window~\cite{DeRocco:2022irl}.} are sensitive to fast mode perturbations from dark matter in the mass range $m \sim 10^{-24} - 10^{-22}\ \mathrm{eV}$ and to slow mode perturbations for masses $m \sim 10^{-18} - 10^{-16}\ \mathrm{eV}$.
These mass ranges coincide with the lower bounds constrained by small-scale structure observations, $m \gtrsim 10^{-22} - 10^{-20}\ \mathrm{eV}$~\cite{Bozek:2014_lowerbound,Irsic:2017_lowerbound,Armengaud:2017_lowerbound,Kobayashi:2017_lowerbound,Zhang:2017_lowerbound,Nori:2018_lowerbound,Leong:2018_lowerbound,Schutz:2020_lowerbound,Rogers:2020_lowerbound,DES:2020_lowerbound,Dalal:2022_lowerbound}. 
Below this threshold, however, ultralight fields can still constitute a subcomponent of dark matter. 

Searches for ultralight dark matter using pulsar timing have been conducted by PPTA~\cite{Porayko:2018_PPTA}, EPTA~\cite{EPTA:2023,EPTA:2023_InPTA}, InPTA~\cite{EPTA:2023_InPTA}, and NANOGrav~\cite{NANOGrav:2023}. 
Current data have achieved sensitivity to dark matter densities around the expected local density ($\simeq 0.3\ \mathrm{GeV/cm^3}$) at $m \sim 10^{-24} ~\mathrm{eV}$ and densities of $\mathcal{O}(10^2) \ \mathrm{GeV/cm^3}$ for $m \sim 10^{-22}\ \mathrm{eV}$.

Vector dark matter, i.e., a spin-$ 1 $ boson, induces metric perturbations distinct from those of scalar dark matter, particularly in the fast mode, with the resulting pulsar timing signals comprising a mixture of scalar, vector, and tensor perturbations at leading order.
This feature leads to notable, distinctive observational signatures for the pulsar timing detection of ultralight vector dark matter~\cite{Nomura:2019_PTA,Unal:2022_PTA,Nomura:2023_PTA}.
Similar signatures have also been investigated in the context of laser interferometers~\cite{Yong:2024_interferometer} and astrometry~\cite{Chen:2022_Astrometry,An:2024_astrometry}. 

However, existing investigations into the pulsar timing detection of vector dark matter remain incomplete. 
Notably, current analyses have solely focused on the fast mode and do not fully capture its statistical properties, with the dark matter field described as a single plane wave at the fast mode frequency $\omega \simeq 2m $ rather than as a stochastic collection of particles, a point emphasized in the context of scalar dark matter in Ref.~\cite{Dror:2025_PTA}. 
The relevant coherence time and coherence length, which govern the correlations of the dark matter field and its associated metric perturbations, are determined by the local velocity dispersion of dark matter, $v_0$:
\begin{align}
    \tau &\equiv \frac{1}{m v_0^2} \simeq 7.4 \times 10^6\, \mathrm{yr}\ \bigg(\frac{m}{10^{-23}\, \mathrm{eV}}\bigg)^{-1} \bigg(\frac{v_0}{160\, \mathrm{km/s}}\bigg)^{-2}, \label{coherence time}\\
    \ell  &\equiv \frac{1}{m v_0} \simeq 1.2\, \mathrm{kpc}\  \bigg(\frac{m}{10^{-23}\, \mathrm{eV}}\bigg)^{-1} \bigg(\frac{v_0}{160\, \mathrm{km/s}}\bigg)^{-1}. \label{coherence length}
\end{align}
The single-plane-wave approximation is valid only when the characteristic time and spatial scales of the problem are much smaller than their respective coherence scales. 
For pulsar timing arrays, Earth-pulsar and pulsar-pulsar separations range from around 100 pc to a few kpc. 
For the fast mode oscillations in the mass range $m \sim 10^{-24} - 10^{-22}\ \mathrm{eV}$, the light travel time is significantly shorter than the coherence time, justifying the neglect of temporal stochasticity. 
However, spatial stochasticity must be accounted for, as the Earth-pulsar and pulsar-pulsar distances are comparable to the coherence length across much of the relevant mass range. 

Current analyses generally adopt one of two limiting approximations: the {\em fully correlated regime} and the {\em fully uncorrelated regime}~\cite{Porayko:2018_PPTA,EPTA:2023,EPTA:2023_InPTA,NANOGrav:2023}.
In the former, the field is treated as a single plane wave with a random phase and amplitude.
In the latter, stochasticity is incorporated by treating the field as a plane wave at the Earth and each pulsar, but with completely uncorrelated phases and amplitudes.

Our work extends beyond the existing literature in several key aspects. 
First, we formulate the local vector dark matter field within a statistical framework, incorporating both the fast and the slow modes with their full statistical properties.
We examine the fast mode with the field stochastic in space while remaining deterministic in time, as performed in Ref.~\cite{Dror:2025_PTA} for scalar dark matter.
We also extend the analysis of vector dark matter to include the slow mode, a topic which has thus far only been explored for scalar dark matter~\cite{Kim:2023_PTA}.
Second, we exhibit the observational signatures via the two-point correlation function of timing residuals, as the metric perturbations induced by the stochastic vector field essentially form a stochastic background.~\footnote{The two-point correlation function of timing residuals for ultralight vector dark matter was explored in Ref.~\cite{Nomura:2023_PTA} and~\cite{Zhu:2024_PTAv} but with the ultralight vector field treated as a plane wave with a single momentum. They hence do not account for the velocity dispersion of the field.} 
Third, we provide a more precise description of the field's polarizations, addressing a limitation of previous studies~\cite{Nomura:2019_PTA,Nomura:2023_PTA} that assumed a linearly polarized single-plane-wave vector field.
Given the nonlinear collapse and virialization of dark matter in the galactic halo, equipartition among the three polarizations is a reasonable assumption, even if production mechanisms may generate an initially polarized state.

The paper is organized as follows. 
In Sec.~\ref{sec2}, we establish the formalism describing the local ultralight dark matter field, demonstrating its wave nature and discussing its coherence and stochasticity on different scales. 
In Sec.~\ref{sec3}, we analyze the structure of the metric perturbations and timing residuals induced by ultralight vector dark matter, laying the basis for the calculation of their two-point correlation functions. 
In Sec.~\ref{sec4}, we derive the explicit expression of the two-point correlation function of timing residuals for the fast mode and discuss its characteristics. 
In Sec.~\ref{sec5}, we show that the slow mode timing residual correlation function for ultralight vector dark matter closely resemble that of scalar dark matter, allowing both to be searched for simultaneously using existing analyses. 
We summarize the paper in Sec.~\ref{sec6}.

We adopt natural units setting $ \hbar = c = 1$ and the metric convention $\eta_{\mu\nu}=\mathrm{diag}(-1,1,1,1)$.

\section{\label{sec2}The Local Ultralight Vector Dark Matter Field}

In this section, we develop a statistical formalism for the local ultralight vector dark matter field, which is essential for capturing its statistical properties in pulsar timing residuals.
We begin with a quantum statistical description of the vector field and demonstrate that, due to its large occupation numbers, the local field behaves classically -- a feature that has been thoroughly addressed only in the case of scalar dark matter~\cite{Rodd:2024_coherent}.
Using this formalism, we derive the relevant field correlators needed for subsequent calculations and examine the behavior of the vector field both within and beyond the coherence scales defined in Eqs.~\eqref{coherence time} and~\eqref{coherence length}.

\subsection{The Quantum Statistical Formalism and Classical Nature}
On the spatial scale of pulsar timing observations (a few kpc), the local ultralight vector dark matter field can be approximated as a free field, with its spatial inhomogeneity due to the gravitational potential of the halo neglected.

The field operator is
\begin{align}
\begin{aligned}
    A_{\mu}(x) = \frac{1}{\sqrt{V}} \sum_{\mathbf{k},\lambda} \frac{1}{\sqrt{2\omega_{\mathbf{k}}}} \Big( a^{\lambda}(\mathbf{k}) \epsilon_{\mu}^{\lambda}(\mathbf{k}) e^{ik \cdot x} + \mathrm{h.c.} \Big),
    \label{field operator}
\end{aligned}
\end{align}
where $k \cdot x = -\omega_{\mathbf{k}}t + \mathbf{k} \cdot \mathbf{x}$ is the inner product of the coordinate and momentum 4-vectors and $\omega_{\mathbf{k}} = \sqrt{|\mathbf{k}|^2 + m^2}$.
$\epsilon_{\mu}^{\lambda}(\mathbf{k})$, $\lambda = +, -, L$ are the three polarization unit vectors for momentum $\mathbf{k}$, satisfying the orthonormality and completeness relations
\begin{align}
    \epsilon_{\mu}^{\lambda}(\mathbf{k}) \epsilon^{\lambda'\mu*}(\mathbf{k}) &= \delta_{\lambda\lambda'}, \\
    \sum_{\lambda} \epsilon_{\mu}^{\lambda}(\mathbf{k}) \epsilon_{\nu}^{\lambda*}(\mathbf{k}) &= \eta_{\mu\nu} + \frac{k_\mu k_\nu}{m^2}. \label{completeness relation}
\end{align}
The annihilation and creation operators for momentum $\mathbf{k}$ and polarization $\lambda$, $a^{\lambda}(\mathbf{k})$ and $a^{\lambda\dagger}(\mathbf{k})$, satisfy the commutation relation
\begin{align}
    \left[a^{\lambda}(\mathbf{k}),a^{\lambda'\dagger}(\mathbf{k}')\right] = \delta_{\mathbf{k},\mathbf{k}'} \delta_{\lambda\lambda'}.
    \label{commutation}
\end{align}
The normalization prefactor $1/\sqrt{V}$, with the volume $V$ defined as
\begin{align}
    V = \int d^3 x\, e^{-i \mathbf{0} \cdot \mathbf{x}} = (2\pi)^3 \delta^3 (0),
\end{align}
originates from replacing $\int d^3 k$ with $\sum_{\mathbf{k}}$ and the usual continuous canonical commutation relation with Eq.~\eqref{commutation}.~\footnote{The two formulations are equivalent, transforming into each other through the rescaling of $a^{\lambda}(\mathbf{k})$. We select the discrete one, Eqs.~\eqref{field operator} and~\eqref{commutation}, for the simplicity and clarity of our subsequent calculations.}

The statistical properties of the local dark matter field are encapsulated in its density operator. It can be expressed in terms of coherent states with the Glauber-Sudarshan $\mathcal{P}$-representation~\cite{Glauber:1963_coherent,Sudarshan:1963_coherent}:
\begin{align}
    \hat{\rho} = \int \Big(\prod_{\mathbf{k},\lambda} d^2 \alpha^{\lambda}(\mathbf{k})\Big)\, \mathcal{P}(\{\alpha^{\lambda}(\mathbf{k})\})  \ket{\{\alpha^{\lambda}(\mathbf{k})\}} \bra{\{\alpha^{\lambda}(\mathbf{k})\}}.
    \label{density operator}
\end{align}
$\ket{\{\alpha^{\lambda}(\mathbf{k})\}}$ is the coherent state, i.e., 
\begin{align}
    a^{\lambda}(\mathbf{k}) \ket{\{\alpha^{\lambda'}(\mathbf{k}')\}} = \alpha^{\lambda}(\mathbf{k})\ket{\{\alpha^{\lambda'}(\mathbf{k}')\}},
\end{align}
with $\alpha^{\lambda}(\mathbf{k})$ the eigenvalue of $a^{\lambda}(\mathbf{k})$.
If each mode of the local dark matter field consists of a superposition of a large number of similar and statistically independent excitations,~\footnote{Such an assumption is plausible provided that the local dark matter system has experienced gravitational collapse and achieved virialization. While this is commonly adopted in the literature, its validity has been questioned and some alternatives were postulated in Ref.~\cite{Rodd:2024_coherent}.}
the weight function $\mathcal{P}(\{\alpha^{\lambda}(\mathbf{k})\})$ is Gaussian distribution~\cite{Glauber:1963_coherent,Rodd:2024_coherent}:
\begin{align}
    \mathcal{P}(\{\alpha^{\lambda}(\mathbf{k})\}) = \prod_{\mathbf{k},\lambda} \frac{1}{\pi N^{\lambda}(\mathbf{k})} e^{-|\alpha^{\lambda}(\mathbf{k})|^2/N^{\lambda}(\mathbf{k})},
    \label{weight function}
\end{align}
where $N^{\lambda}(\mathbf{k}) \equiv \left< |\alpha^{\lambda}(\mathbf{k})|^2 \right>$ is the expectation value of the occupation number of the mode $(\mathbf{k},\lambda)$. 
This distribution of particles over the modes is determined by the phase space probability distribution of the dark matter particles, $P(\mathbf{k},\lambda)$, or the phase space probability density, $f(\mathbf{k},\lambda)$, as~\footnote{We use $\mathcal{P}$ for the Glauber-Sudarshan $\mathcal{P}$-representation, $P$ as probability, and $f$ as probability density.}
\begin{align}
    N^{\lambda}(\mathbf{k}) = N P(\mathbf{k},\lambda) = N f(\mathbf{k},\lambda) \,d^3 k,
\end{align}
with $N$ the total particle number in the system.
Eq.~\eqref{weight function} implies that the eigenvalues $\alpha^{\lambda}(\mathbf{k})$ for different modes are statistically independent complex Gaussian random variables. 

The classical nature of the local ultralight vector dark matter field is a consequence of its large state occupation numbers.
The quantum fluctuations of any operator $O$ associated with the field on any coherent state $\ket{\{\alpha\}} \equiv \ket{\{\alpha^{\lambda}(\mathbf{k})\}}$:
\begin{align}
    \delta O |_{\{\alpha\}} \equiv \frac{\sqrt{\left< \{\alpha\}| O^2 | \{\alpha\} \right> - \left< \{\alpha\}| O | \{\alpha\} \right>^2}}{\left< \{\alpha\}| O | \{\alpha\} \right>},
\end{align}
acquires a nonzero value from the commutators, Eq.~\eqref{commutation}, and vanishes when the commutators are taken to be zero.
In the large occupation number limit,
\begin{align}\label{large occupation number}
    \left[a^{\lambda}(\mathbf{k}),a^{\lambda\dagger}(\mathbf{k})\right] = 1 \ll N^{\lambda}(\mathbf{k}),
\end{align}
the quantum fluctuations are $\mathcal{O}(1/\sqrt{N})$ and thus negligible.
The vector field essentially behaves as a classical field.

For the field operator, Eq.~\eqref{field operator}, with commutators neglected, $a^{\lambda}(\mathbf{k})$ reduces to its $c$-number eigenvalues $\alpha^{\lambda}(\mathbf{k})$.
It consequently turns into a {\em stochastic} classical field specified as a sum of classical plane waves of random amplitudes and phases:
\begin{align}
\begin{aligned}
    A_{\mu}(x) = \frac{1}{\sqrt{V}} \sum_{\mathbf{k},\lambda} \frac{1}{\sqrt{2\omega_{\mathbf{k}}}} \Big( \alpha^{\lambda}(\mathbf{k}) \epsilon_{\mu}^{\lambda}(\mathbf{k}) e^{ik \cdot x} + \mathrm{c.c.} \Big),
    \label{classical field}
\end{aligned}
\end{align}
with the random complex variables $\{\alpha^{\lambda}(\mathbf{k})\}$ generated by Eq.~\eqref{weight function}.

In the quantum statistical formalism, the expectation value of any operator $O$ is calculated with the density operator as
\begin{align}\label{expectation}
    &\left< O \right> = \mathrm{Tr} (O \hat{\rho}) \\
    &= \int \Big(\prod_{\mathbf{k},\lambda} d^2 \alpha^{\lambda}(\mathbf{k})\Big) \mathcal{P}(\{\alpha^{\lambda}(\mathbf{k})\}) \left< \{\alpha^{\lambda}(\mathbf{k})\}| O | \{\alpha^{\lambda}(\mathbf{k})\} \right>. \notag
\end{align}
It can be approximated by the expectation value of the corresponding quantity associated with the classical field, with $a^{\lambda}(\mathbf{k})$ and $a^{\lambda\dagger}(\mathbf{k})$ replaced by their eigenvalues $\alpha^{\lambda}(\mathbf{k})$ and $\alpha^{\lambda*}(\mathbf{k})$, as the ordering of $a^{\lambda}(\mathbf{k})$ and $a^{\lambda\dagger}(\mathbf{k})$ only introduces a $\mathcal{O}\left(1/N^{\lambda}(\mathbf{k})\right)$ correction to the expectation value.
This yields the correspondence:
\begin{align}
\begin{aligned}
    &\left< O(\{ a^{\lambda}(\mathbf{k}) \}, \{ a^{\lambda\dagger}(\mathbf{k}) \}) \right> \simeq \left< O(\{ \alpha^{\lambda}(\mathbf{k}) \}, \{ \alpha^{\lambda*}(\mathbf{k}) \}) \right> \\
    &= \int \Big(\prod_{\mathbf{k},\lambda} d^2 \alpha^{\lambda}(\mathbf{k})\Big) \mathcal{P}(\{\alpha^{\lambda}(\mathbf{k})\}) O(\{ \alpha^{\lambda}(\mathbf{k}) \}, \{ \alpha^{\lambda*}(\mathbf{k}) \}).
\end{aligned}
\end{align}

For our subsequent calculations, we present the two-point and four-point correlators of $\alpha^{\lambda}(\mathbf{k})$ and $\alpha^{\lambda*}(\mathbf{k})$. 
The only nonzero two-point correlator is 
\begin{align}
    \left< \alpha^{\lambda*}(\mathbf{k}) \alpha^{\lambda'}(\mathbf{k}') \right> = N P(\mathbf{k},\lambda) \delta_{\mathbf{k},\mathbf{k'}} \delta_{\lambda\lambda'},
\end{align}
and the only nonzero four-point correlator is
\begin{align}\label{4-point a adagger correlator}
    &\left< \alpha^{\lambda_1*}(\mathbf{k}_1) \alpha^{\lambda_2*}(\mathbf{k}_2) \alpha^{\lambda_3}(\mathbf{k}_3) \alpha^{\lambda_4}(\mathbf{k}_4) \right> \\
    &= N^2 P(\mathbf{k}_1,\lambda_1) P(\mathbf{k}_2,\lambda_2) \big( \delta_{\mathbf{k}_1,\mathbf{k}_3} \delta_{\lambda_1\lambda_3} \delta_{\mathbf{k}_3,\mathbf{k}_4} \delta_{\lambda_2\lambda_4} \notag \\
    & \hphantom{= N^2 P(\mathbf{k}_1,\lambda_1) P(\mathbf{k}_2,\lambda_2)}+ \delta_{\mathbf{k}_1,\mathbf{k}_4} \delta_{\lambda_1\lambda_4} \delta_{\mathbf{k}_2,\mathbf{k}_3} \delta_{\lambda_2\lambda_3} \big). \notag
\end{align}

\subsection{Behavior and Coherence Scales}

We now examine the behavior of the stochastic classical field, Eq.~\eqref{classical field}, under the equipartition assumption, $P(\mathbf{k},\lambda) = P(\mathbf{k})P(\lambda) =  P(\mathbf{k})/3$, in the non-relativistic limit.
In this case, the polarizations for each momentum $\mathbf{k}$ can be decomposed onto a common basis $\left\{\hat{x},\hat{y},\hat{z}\right\}$ with particles uniformly distributed on them.
The classical field is therefore:
\begin{align}
\begin{aligned}
    A_{i}(x) = \frac{1}{\sqrt{2mV}} \sum_{\mathbf{k}} \Big( \alpha^{i}(\mathbf{k}) e^{ik \cdot x} + \mathrm{c.c.} \Big),
    \label{non-relativistic classical field}
\end{aligned}
\end{align}
where the $i$ in $\alpha^{i}(\mathbf{k})$ now represents the polarization along the $i$-th axis, and the 0-th component of the field is determined through the condition $\partial_\mu A^\mu = 0$.

The behavior of this sum of plane waves can be characterized by its coherence scales defined in Eqs.~\eqref{coherence time} and~\eqref{coherence length}. 
Well below the coherence scales, the field exhibits coherent oscillation and is approximately described by a single plane wave,
\begin{align}
\begin{aligned}
     A_{i}(x) &\simeq \frac{1}{2} \frac{\sqrt{2\bar{\rho}}}{m} \Big( \beta_{i} e^{ip \cdot x} +  \mathrm{c.c.} \Big) \\
     &= \frac{\sqrt{2\bar{\rho}}}{m} |\beta_i| \cos \left( \omega_{\mathbf{p}}t - \mathbf{p} \cdot \mathbf{x} - \varphi_i \right),
     \label{coherent classical field}
\end{aligned}
\end{align}
with a random complex vector $\beta_i = |\beta_i|\, e^{i\varphi_i}$ that encodes the statistics of its amplitude and polarization, normalized with the mean local dark matter density $\bar{\rho}$, and a randomly drawn wavevector $\mathbf{p}$. 
The two descriptions Eqs.~\eqref{non-relativistic classical field} and \eqref{coherent classical field} must agree with each other at least locally at one spacetime point. 
Equating Eqs.~\eqref{non-relativistic classical field} and \eqref{coherent classical field} and their time derivatives at $x^{\mu} = 0$, we obtain the expression of $\beta_i$:
\begin{align}
\begin{aligned}
    \beta_i = \sqrt{\frac{m}{V\bar{\rho}}} \sum_{\mathbf{k}} \alpha^{i}(\mathbf{k}).
    \label{beta_i}
\end{aligned}
\end{align}
In the summation over $\mathbf{k}$, each Gaussian random variable $\alpha^{i}(\mathbf{k})$ can be broken down into a sum of $\sqrt{N^{i}(\mathbf{k})}$ independent random variables generated by the same Gaussian distribution with variance equal to 1.
Thus, the summation in Eq.~\eqref{beta_i} can be rewritten as a sum over $\sum_{\mathbf{k}} \sqrt{N^{i}(\mathbf{k})}$ number of such independent and identically distributed variables, which, with the central limit theorem, implies that the vector $\beta_i$ also follows a Gaussian distribution:
\begin{align}
    f(\{\beta_i\}) = \prod_{i=1}^3 \frac{3}{\pi} e^{-3|\beta_i|^2}.
    \label{beta_i distribution}
\end{align}
Although the expectation value of the squared norm $\left< \sum_i |\beta_i|^2 \right> = 1$, $\beta_i$ is not generically a unit vector. 
This isotropic, $\varphi_i$-independent distribution indicates that the polarization of the plane wave field, Eq.~\eqref{coherent classical field}, is generically elliptical rather than linear~\cite{Amin:2024_equipartition}.

However, close to or beyond the coherence scales, the field starts to exhibit stochasticity.
This can be shown with the two-point field correlation function.
Applying the non-relativistic limit and assuming equipartition,
\begin{align}
\begin{aligned}
    \left< A_\mu(x) A_\nu(x') \right> = \frac{N \eta_{\mu\nu}}{3mV} \sum_{\mathbf{k}} P(\mathbf{k}) \cos (k \cdot \Delta x),
\end{aligned}
\end{align}
where $k \cdot \Delta x = - \omega_\mathbf{k} \Delta t + \mathbf{k} \cdot \Delta \mathbf{x}$, with $\omega_\mathbf{k} \simeq m + |\mathbf{k}|^2/2m$, $\Delta t \equiv t-t'$ and $\Delta \mathbf{x} \equiv \mathbf{x}-\mathbf{x}'$.
When the space and time separations are well below the coherence scales implicitly contained in $P({\bf k})$ ($|\Delta t| \ll \tau$ and $|\Delta \mathbf{x}| \ll \ell$), the cosines in the sum constructively interfere maximizing the correlation.
When the space and time separations exceed the coherence scales ($|\Delta t| \gtrsim \tau$ or $|\Delta \mathbf{x}| \gtrsim \ell$), the cosine completes full cycles within the width of $P(\mathbf{k})$ and the summation over cosines partially cancel out, reducing the correlation. In this limit, the single-plane-wave description in Eq.~\eqref{coherent classical field} breaks down. 
To fully capture the statistical properties of ultralight vector dark matter in this regime, one must adopt the field description as a superposition of plane waves, Eq.~\eqref{classical field} or \eqref{non-relativistic classical field}. 

In our following discussions, we employ the classical field description, Eq.~\eqref{classical field}, in the non-relativistic limit. 
To keep our results widely applicable, we refrain from assuming equipartition until the phase space summations are explicitly performed. 

\section{\label{sec3}Framework of Pulsar Timing Signals}

The principal observable relevant to this work is the two-point correlation function of the timing residuals.
The timing residuals arise from the metric perturbations, which are sourced by the energy-momentum tensor of the local vector dark matter field.
Therefore, in this section, we construct the framework connecting the local vector dark matter field to the timing residuals.
In order to identify the leading-order contributions and the characteristics of the fast and the slow mode signals, we perform a structural analysis of the energy-momentum tensor, metric perturbations, and timing residuals.
This approach is powerful, as it reveals the most convenient way to understand the influence of each mode on a pulsar timing experiment at the leading order in the velocity expansion.

\subsection{Structure of Metric Perturbations}

In the weak-field limit, the spacetime metric $g_{\mu\nu}$ can be expanded into the Minkowski flat background and a small perturbation $h_{\mu\nu}$:
\begin{align}
    g_{\mu\nu} = \eta_{\mu\nu} + h_{\mu\nu}\, ,\ \ |h_{\mu\nu}|\ll1.
\end{align}
$h_{\mu\nu}$ can be further decomposed into several components~\cite{Carroll:2004st}:
\begin{align}
    \left\{
    \begin{aligned}
        h_{00} &= -2\Phi \\
        h_{0i} &= w_i \\
        h_{ij} &= 2s_{ij} -2\Psi \delta_{ij}
    \end{aligned}
    \right.
    \; ,
\end{align}
with 
\begin{align}
\begin{aligned}
    \Psi &= -\frac{1}{6} \delta_{ij} h_{ij}, \\
    s_{ij} &= \frac{1}{2} \left( h_{ij} - \frac{1}{3} \delta_{kl} h_{kl} \delta_{ij}, \right), \\
\end{aligned}
\end{align}
encoding the trace and the traceless part of $h_{ij}$ respectively.
These metric perturbations are sourced by the energy-momentum tensor through the Einstein's equations, which can be solved after fixing the gauge degrees of freedom.

In this paper, we work in two commonly adopted gauges.
One is the {\em transverse gauge}, where
\begin{align}
    \partial_i w_i = \partial_i s_{ij} = 0.
\end{align}
This gauge condition eliminates the scalar mode in $w_i$, and the scalar and the vector mode in $s_{ij}$, leaving $w_i$ purely a vector perturbation and $s_{ij}$ purely a tensor perturbation.
It is thus an effective means to identify the individual contributions of the scalar, vector, and tensor perturbations.

The other is the {\em synchronous gauge}, where
\begin{align}
    \Phi = w_i =0.
\end{align}
In this gauge, most components of metric perturbations are eliminated and calculations are simplified when multiple components are involved. 
We provide the full solutions to the Einstein's equations in these two gauges in App.~\ref{apdx1}.~\footnote{\label{inverse operator}In this paper, we use $\frac{1}{\partial_0}$, $\frac{1}{\nabla^2}$, and $\frac{1}{\Box}$ to denote the inverse operation of the differential operators $\partial_0$, $\nabla^2$, and $\Box$ for the sake of succinctness of our formulas, i.e., for any function of spacetime $f(t,\mathbf{x})$:
\begin{align}
    \frac{1}{\partial_0}f(t,\mathbf{x}) &= \int^t dt'\, f(t',\mathbf{x}),\\
    \frac{1}{\nabla^2}f(t,\mathbf{x}) &= - \frac{1}{4\pi} \int d^3x'\, \frac{1}{|\mathbf{x}-\mathbf{x}'|} f(t,\mathbf{x}'),\\
    \frac{1}{\Box}f(t,\mathbf{x}) &= - \frac{1}{4\pi} \int d^4x'\, \frac{\delta(t-t'- |\mathbf{x}-\mathbf{x}'|)}{|\mathbf{x}-\mathbf{x}'|} f(t',\mathbf{x}').
\end{align}
The boundary conditions are not observable with pulsar timing arrays, so we can safely discard them.
}

To derive the structure of the metric perturbations, we first look into the energy-momentum tensor of ultralight vector dark matter:
\begin{align}\label{energy-momentum tensor}
\begin{aligned}
    T_{\mu\nu} ={}& F_{\mu\rho} \tensor{F}{_\nu^\rho} - \frac{1}{4} \eta_{\mu\nu} F_{\rho\sigma} F^{\rho\sigma} \\
    &+ m^2 \left( A_\mu A_\nu - \frac{1}{2} \eta_{\mu\nu} A_\rho A^\rho \right).
\end{aligned}
\end{align}
As $T_{\mu\nu}$ is quadratic in $A_\mu$, which is itself a sum of multiple plane waves, it consists of two characteristic modes: 
\begin{align}
    T_{\mu\nu} = \sum_{\mathbf{k},\lambda} \sum_{\mathbf{k}',\lambda'} \Big( \mathcal{F}_{\mu\nu} e^{i (k+k') \cdot x} + \mathcal{S}_{\mu\nu} e^{i (k-k') \cdot x} + \mathrm{c.c.} \Big),
\end{align}
where $\mathcal{F}_{\mu\nu}$ and $\mathcal{S}_{\mu\nu}$ depend on $\mathbf{k}$, $\lambda$, $\mathbf{k}'$, and $\lambda'$.
The $\mathcal{F}_{\mu\nu}$ terms, which we refer to as the {\em fast mode}, are characterized by the time dependence $e^{ i(\omega_{\mathbf{k}}+\omega_{\mathbf{k}'})t}$, with angular frequencies $\omega_{\mathbf{k}} + \omega_{\mathbf{k}'} \simeq 2m$. 
The $\mathcal{S}_{\mu\nu}$ terms, which we refer to as the {\em slow mode}, are characterized by the time dependence $e^{i(\omega_{\mathbf{k}}-\omega_{\mathbf{k}'})t}$, with angular frequencies $\omega_{\mathbf{k}} - \omega_{\mathbf{k}'} \simeq (|\mathbf{k}|^2-|\mathbf{k}'|^2)/2m$,  which are $\mathcal{O}(v_0^2) \sim 10^{-6}$ smaller than the fast mode frequencies.
Spatially, both the fast and the slow modes have de Broglie-scale wavelengths, with their wavenumbers $|\mathbf{k}\pm\mathbf{k}'|\sim mv_0$.

Physically, the fast mode corresponds to the coherent oscillation of the field. 
It has a vanishing expectation value, as $\langle \mathcal{F}_{\mu\nu} \rangle$ contains only $\langle \alpha^{\lambda}(\mathbf{k})\alpha^{\lambda'}(\mathbf{k}') \rangle$ or $\langle \alpha^{\lambda*}(\mathbf{k})\alpha^{\lambda'*}(\mathbf{k}') \rangle$ that vanish. In the single-plane-wave approximation for the field, Eq.~\eqref{coherent classical field}, the wavevectors $k+k' \rightarrow 2p$, and the fast mode reduces to a single oscillating mode in the energy-momentum tensor. 

In contrast, the slow mode is a manifestation of the interference pattern of the field.
The expectation value of the slow mode does not always vanish; $\langle \mathcal{S}_{\mu\nu} \rangle$ is composed of $\langle \alpha^{\lambda}(\mathbf{k})\alpha^{\lambda'*}(\mathbf{k}') \rangle$, which contributes when $k=k'$, eliminating the spacetime dependence $e^{i(k-k')\cdot x}$. 
Particularly, the expectation value of $T_{00}$ is the mean local dark matter density $\bar{\rho}$:
\begin{align}\label{mean local dark matter density}
    \bar{\rho} = \left< T_{00} \right> \simeq \frac{mN}{V}.
\end{align}
In the single-plane-wave approximation, $k-k' \rightarrow 0$, interference disappears and the slow mode is reduced to a constant, characterizing the mean properties of the field.

\newcolumntype{C}{>{\centering\arraybackslash}X}
\newcolumntype{D}{>{$\displaystyle}C<{$}}
\renewcommand{\arraystretch}{1.75}
\begin{table*}
\centering
\caption{\label{parametric sizes}Structures of the energy-momentum tensors and metric perturbations of ultralight scalar and vector dark matter, expressed in terms of the parametric sizes of the fast and slow modes of each component, where $\bar{\rho}$ denotes the mean local dark matter density, $v_0 \sim 10^{-3}$ represents the typical velocity of local dark matter, and $G$ is the gravitational constant. 
The parametric sizes of $P_{ij} T_{0j}$ and $\Lambda_{ij,kl} T_{kl}$, where $P_{ij} = \delta_{ij} - \partial_i \partial_j /\nabla^2$ is the transverse projector and $\Lambda_{ij,kl} = P_{ik} P_{jl} - P_{ij} P_{kl} /2$ is the transverse-traceless projector, are also provided, as they are directly related to $w_i$ and $s_{ij}$ in the transverse gauge.
The metric perturbations in red boxes are leading-order contributors to the corresponding mode timing residuals.}
\begin{tabularx}{\linewidth}{c|c|DDDDDD|DDDD|DD}
    \toprule
    \toprule
    \multicolumn{2}{c|}{\multirow{3}{*}{}} & \multicolumn{6}{c|}{\multirow{2}{*}{\textrm{Energy-momentum tensor $\displaystyle \left/ \bar{\rho} \right.$}}} & \multicolumn{6}{c}{Metric perturbations $\displaystyle \left/ \frac{G\bar{\rho}}{m^2} \right.$} \\[5pt] \cline{9-14}
    \multicolumn{2}{c|}{} & \multicolumn{6}{c|}{} & \multicolumn{4}{c|}{Transverse gauge} & \multicolumn{2}{c}{Synchronous gauge} \\ 
    \cline{3-14}
    \multicolumn{2}{c|}{} & T_{00} & T_{0i} & T_{kk} & \pi_{ij} & P_{ij}T_{0j} & \Lambda_{ij,kl} T_{kl} & \Psi & \Phi & w_i & s_{ij} & \Psi & s_{ij} \\
    \hline
    \rule{0pt}{4ex}
    \multirow{2}{*}{\rotatebox{90}{ULSDM }} & Fast mode & v_0^2 & v_0 & 1 & v_0^2 & v_0^3~\footnote{The leading-order term of $T_{0i}$ of scalar dark matter ($\mathcal{O}(\bar{\rho}v_0)$) is eliminated by the transverse projection tensor $P_{ij}$, which does not occur in the case of vector dark matter.} & v_0^2 & \fcolorbox{red}{white}{$\displaystyle 1$} & 1 & v_0 & v_0^2 & \fcolorbox{red}{white}{$\displaystyle 1$} & v_0^2 \\[3pt]
    & Slow mode & 1 & v_0 & v_0^2 & v_0^2 & v_0 & v_0^2 & v_0^{-2} & \fcolorbox{red}{white}{$\displaystyle v_0^{-2}$} & v_0^{-1} & 1 & \fcolorbox{red}{white}{$\displaystyle v_0^{-4}$} & \fcolorbox{red}{white}{$\displaystyle v_0^{-4}$} \\[3pt]
    \hline
    \rule{0pt}{4ex}
    \multirow{2}{*}{\rotatebox{90}{ULVDM }} & Fast mode & v_0^2 & v_0 & 1 & 1 & v_0 & 1 & \fcolorbox{red}{white}{$\displaystyle 1$} & \fcolorbox{red}{white}{$\displaystyle v_0^{-2}$} & \fcolorbox{red}{white}{$\displaystyle v_0^{-1}$} & \fcolorbox{red}{white}{$\displaystyle 1$} & \fcolorbox{red}{white}{$\displaystyle 1$} & \fcolorbox{red}{white}{$\displaystyle 1$} \\[3pt]
    & Slow mode & 1 & v_0 & v_0^2 & v_0^2 & v_0 & v_0^2 & v_0^{-2} & \fcolorbox{red}{white}{$\displaystyle v_0^{-2}$} & v_0^{-1} & 1 & \fcolorbox{red}{white}{$\displaystyle v_0^{-4}$} & \fcolorbox{red}{white}{$\displaystyle v_0^{-4}$} \\[3pt]
    \bottomrule
    \bottomrule
\end{tabularx}
\end{table*}

The parametric size of each component of the fast and the slow modes of $T_{\mu\nu}$ in the velocity expansion can be derived by plugging the field, Eq.~\eqref{field operator}, into Eq.~\eqref{energy-momentum tensor}.
The results in terms of components $T_{00}$, $T_{0i}$, $T_{kk}$, and $\pi_{ij} \equiv T_{ij} - T_{kk}\delta_{ij}/3$ for each mode are summarized on the left side of Tab.~\ref{parametric sizes}.
We also present the parametric sizes of the transversely projected components, $P_{ij} T_{0j}$, and the transverse-traceless projected component, $\Lambda_{ij,kl} T_{kl}$, in Tab.~\ref{parametric sizes}, as they are necessary for inferring the parametric sizes of the metric perturbations in the transverse gauge.

The fast and the slow modes of $T_{\mu\nu}$ generate the fast and the slow mode metric perturbations respectively in the weak-field limit.
Acting on these two modes, the differential operators in the Einstein's equations produce factors of different parametric sizes:
\begin{align}
\begin{aligned}
    \text{Fast mode}:&\ \
    \begin{aligned}
        \partial_0 \sim m,\ \partial_i \sim mv_0,\ \Box \sim m^2;
    \end{aligned}\\
    \text{Slow mode}:&\ \
    \begin{aligned}
        \partial_0 \sim mv_0^2,\ \partial_i \sim mv_0,\ \Box \sim m^2v_0^2.
    \end{aligned}
\end{aligned}
\label{sizes of differential operators}
\end{align}
Then, the parametric sizes of metric perturbations $\Psi$, $\Phi$, $w_i$, and $s_{ij}$ for the fast and the slow modes can be derived in both the transverse and the synchronous gauge, by substituting the parametric sizes of $T_{\mu\nu}$ in Tab.~\ref{parametric sizes} and the scaling, Eq.~\eqref{sizes of differential operators}, into the solutions to Einstein's equations provided in App.~\ref{apdx1}.
We summarize the results on the right side of Tab.~\ref{parametric sizes}. 
In combination, these parametric sizes of metric perturbations in the velocity expansion determine the structure of pulsar timing signals of ultralight vector dark matter in the following discussions.

\subsection{Structure of Timing Residuals and Observability}

The metric perturbations modify the pulsar timing signal through redshifting the received emission frequency.
The redshift, to the first order of $h_{\mu\nu}$, is
\begin{align}
\begin{aligned}
    z ={}& - \hat{n}_i \hat{n}_j \int_{t_{\mathrm{e}}}^{t_{\mathrm{o}}} dt\, \left[ \frac{1}{\partial_0} R_{i0j0} \right]_ {\mathbf{x}=\mathbf{x}_\gamma(t)},
    \label{redshift}
\end{aligned}
\end{align}
where $R_{i0j0}$ is the $i0j0$ component of the Riemann curvature tensor, which is independent of gauge choices and is expressed in terms of metric perturbations by: 
\begin{align}\label{Riemann tensor}
    R_{i0j0} &= \partial_i \partial_j \Phi + \ddot{\Psi} \delta_{ij} +  \frac{1}{2}(\partial_{i} \dot{w}_{j} + \partial_{j} \dot{w}_{i}) - \ddot{s}_{ij}, 
\end{align}
where the dots denote time derivatives, $\partial_0$. 
The unit vector $\hat{\mathbf{n}}$ points backward along the propagation path of the photon, $t_{\mathrm{o}}$ and $t_{\mathrm{e}}$ are the time of observation and emission, and $\mathbf{x}_\gamma(t) = \mathbf{x}_{\mathrm{O}}(t_{\mathrm{o}}) + \left(t_{\mathrm{o}} - t\right) \hat{\mathbf{n}} = \mathbf{x}_{\mathrm{P}}(t_{\mathrm{e}}) + \left(t_{\mathrm{e}} - t\right) \hat{\mathbf{n}}$ is the flat spacetime path of the emitted photon from the pulsar to the observer, with $\mathbf{x}_{\mathrm{O}}(t)$ and $\mathbf{x}_{\mathrm{P}}(t)$ the trajectories of the observer and the pulsar.
The derivation of Eq.~\eqref{redshift} is provided in App.~\ref{apdx2} and we work in a frame where both the observer and pulsar are non-relativistic.
Note that since $R_{i0j0}$ is first order in $h_{\mu\nu}$, all other elements in Eq.~\eqref{redshift} take their zeroth-order values.
The pulsar timing residual is obtained by integrating the redshift over the arrival time:
\begin{align}\label{timing residual}
    r(t) = \int_0^t dt'\, z(t').
\end{align}

Eqs.~\eqref{redshift} and~\eqref{timing residual} indicate that the contribution of each component of metric perturbations are weighted by its contribution to $R_{i0j0}$.
Inserting the parametric sizes of the metric perturbations in Tab.~\ref{parametric sizes} and the scaling of the differential operators, Eq.~\eqref{sizes of differential operators}, into the expression of $R_{i0j0}$, Eq.~\eqref{Riemann tensor}, we can determine the parametric size of the contribution to $R_{i0j0}$ sourced by each component of the metric perturbations and identify the leading-order contributors, which are also the leading-order contributors to the timing residuals, for both the fast and the slow modes.
We highlight them with red boxes in Tab.~\ref{parametric sizes}.
For comparison, we also present the parametric sizes of the energy-momentum tensor and metric perturbations of scalar dark matter and identify the leading-order contributors with red boxes.

The most significant difference between scalar and vector dark matter arises in the fast mode.
This distinction is clearest in the transverse gauge, where the leading-order pulsar timing residuals of vector dark matter arise from a combination of two scalar perturbations, a vector perturbation, and a tensor perturbation.
In contrast, the leading contribution for scalar dark matter arises solely from the scalar perturbation $\Psi$.

For the slow mode, the parametric structures of $T_{\mu\nu}$ and the metric perturbations of vector dark matter mirror those of scalar dark matter.
In the transverse gauge, the leading-order timing residuals for both scalar and vector dark matter are sourced by the scalar perturbation $\Phi$ of the same parametric size.

Thus, we can expect the fast mode timing residuals to offer the most distinctive observational signature of vector dark matter, while the slow mode timing residuals should resemble those of scalar dark matter in both form and amplitude.

Eqs.~\eqref{redshift},~\eqref{Riemann tensor} and~\eqref{timing residual} also allow us to derive the parametric sizes of the fast mode and slow mode timing residuals.
The line-of-sight integral in Eq.~\eqref{redshift} can be considered as the inverse of the directional derivative along the photon path, $(\partial_0 - \hat{n}_i \partial_i)^{-1}$.
In the velocity expansion, according to the scaling, Eq.~\eqref{sizes of differential operators}, it is dominated by $\partial_0^{-1}$ when acting on the fast mode and is dominated by $-(\hat{n}_i \partial_i)^{-1}$ when acting on the slow mode.
It hence produces a factor of $1/m$ for the fast mode and $1/mv_0$ for the slow mode, inverse to those produced by the directional derivative.
The integration over arrival time in Eq.~\eqref{timing residual} contributes to a factor of $\partial_0^{-1} \sim 1/m$ (fast mode) or $1/mv_0^2$ (slow mode).

Thus, the characteristic magnitude of the timing residuals for each mode is:
\begin{align}\label{magnitude of timing residuals}
    r(t) \sim \frac{G\bar{\rho}}{m^3} \left\{
    \begin{array}{cc}
        \displaystyle 1 & (\text{fast mode}) \\
        \displaystyle v_0^{-5} & (\text{slow mode})
    \end{array}
    \right. .
\end{align}
This parametric structure will be shown explicitly with the expressions of the fast mode and the slow mode timing residual two-point correlation functions derived in the following two sections.

The comparison between the fast and the slow mode amplitudes must be interpreted with care.
The fast mode frequency is $\mathcal{O}(v_0^{-2}) \sim 10^6$ times higher than that of the slow mode, so the peak sensitivity of pulsar timing arrays to fast mode signals occurs at masses about $10^{-6}$ lower than for the slow mode.
At fixed local dark matter density, the $1/m^3$ scaling in front implies that the fast mode signals are $10^3$ larger than slow mode when evaluated within the respective pulsar timing frequency bands. 
These scaling relations --- and in particular, Eq.~\eqref{magnitude of timing residuals} --- also apply to scalar dark matter.

The ultimate goal in this work is to calculate two-point correlation function of timing residuals:
\begin{align}\label{residual correlation function}
    \left< r_a(t) r_b(t') \right> = \int_0^t d\tau\, \int_0^{t'} d\tau'\,  \left< z_a(\tau) z_b(\tau') \right>,
\end{align}
where $a$ and $b$ signify pulsar $a$ and pulsar $b$.
Since the fast and slow modes enter with different combinations of the random variables --- fast modes with $\alpha^{\lambda}(\mathbf{k})\alpha^{\lambda'}(\mathbf{k}')$ or $\alpha^{\lambda*}(\mathbf{k})\alpha^{\lambda'*}(\mathbf{k}')$, and slow modes with $\alpha^{\lambda}(\mathbf{k})\alpha^{\lambda'*}(\mathbf{k}')$ --- their cross terms vanish, and the correlation function naturally separates into a fast mode and a slow mode component. In the next two sections, we derive explicit expressions for these components, making use of the leading-order structure outlined above. 

\section{\label{sec4}Fast mode timing residual two-point correlation function}

We now study the fast mode pulsar timing signals of vector dark matter. 
We treat the pulsars as comoving with the solar system and neglect the motion of the Earth around the Sun.~\footnote{The relative motion between the pulsars and the observer has two effects. One is to contribute to the redshift, by both causing the ordinary Doppler redshift and producing a correction of $\mathcal{O}(vh_{\mu\nu})$ to the redshift sourced by metric perturbations, with $v$ the relative velocity (see App.~\ref{apdx2}). The other is to change the time dependence of the timing residuals as the positions of the pulsars relative to the observer are varying with time. This effect turns out to be negligible for both fast and slow modes. We discuss these effects in detail in App.~\ref{apdx3}.}
We work in the solar system rest frame, with the Earth at the origin and the pulsar locations held fixed; i.e., in Eq.~\eqref{redshift}, $\mathbf{x}_\mathrm{O}(t) = \mathbf{0}$ while the pulsar $a$ is at $\mathbf{x}_\mathrm{P}(t) = d_a \hat{\mathbf{n}}_a$, and $t_\mathrm{e} = t_\mathrm{o} - d_a$.

As multiple components of metric perturbations are involved in the fast mode signal of vector dark matter, we evaluate the timing residual two-point correlation function in the {\em synchronous gauge}, where we only need to consider the $\Psi$ and $s_{ij}$ perturbations, both of which contribute at the leading order in the velocity expansion per Tab.~\ref{parametric sizes}. 
Retaining only $\Psi$ and $s_{ij}$ in $R_{i0j0}$ and the leading term in the velocity expansion of the integration along the photon path in Eq.~\eqref{redshift} for the fast mode, $(\partial_0 - \hat{n}_i \partial_i)^{-1} \simeq \partial_0^{-1}$, we obtain the fast mode redshift at arrival time $t$:
\begin{align}\label{fast mode redshift}
\begin{aligned}
    z_a(t) ={}& - \Big( \Psi(t,\mathbf{0}) - \Psi(t-d_a,\ d_a\hat{\mathbf{n}}_a) \Big) \\
    &+ \hat{n}_a^i \hat{n}_a^j \Big( s_{ij}(t,\mathbf{0}) - s_{ij}(t-d_a,\ d_a\hat{\mathbf{n}}_a) \Big).
\end{aligned}
\end{align}

With the structure of the fast mode of $T_{\mu\nu}$ presented in Tab.~\ref{parametric sizes} and the scaling of Eq.~\eqref{sizes of differential operators}, we can identify the leading-order terms in the solutions to the Einstein's equations in the synchronous gauge provided in App.~\ref{apdx1} and obtain the simplified solutions:
\begin{align}
    \Psi &= \frac{4}{3} \pi G \frac{1}{\partial_0^2} T_{kk}\, , \label{simp Psi}\\
    s_{ij} &= -8\pi G \frac{1}{\Box} \pi_{ij}\, \label{simp sij}.
\end{align}
The fast mode term of $T_{ij}$ to the leading order is:
\begin{align}
    T_{ij} ={}& - \dot{A}_i \dot{A}_j + m^2 A_i A_j - \frac{1}{2} \eta_{ij} \Big( - |\dot{\mathbf{A}}|^2 + m^2 |\mathbf{A}|^2 \Big) \notag \\[3pt]
    ={}&  \frac{m}{V} \sum_{\mathbf{k},\lambda} \sum_{\mathbf{k}',\lambda'} \alpha^{\lambda}(\mathbf{k}) \alpha^{\lambda'}(\mathbf{k}') e^{i(k+k') \cdot x} \\ 
    &\times \Big( \epsilon_{i}^{\lambda}(\mathbf{k}) \epsilon_{j}^{\lambda'}(\mathbf{k}') - \frac{1}{2} \delta_{ij} \epsilon_{k}^{\lambda}(\mathbf{k}) \epsilon_{k}^{\lambda'}(\mathbf{k}') \Big)+ \mathrm{c.c.}\ . \notag
\end{align}
Plugging it into Eqs.~\eqref{simp Psi} and~\eqref{simp sij}, we obtain
\begin{align}
&\begin{aligned}
    \Psi ={} \frac{\pi G}{6mV} \sum_{\mathbf{k},\lambda} \sum_{\mathbf{k}',\lambda'} \alpha^{\lambda}(\mathbf{k}) \alpha^{\lambda'}(\mathbf{k}') e^{i(k+k') \cdot x} \\
    \times\ \epsilon_{i}^{\lambda}(\mathbf{k}) \epsilon_{i}^{\lambda'}(\mathbf{k}') + \mathrm{c.c.}\ ,
\end{aligned}\\[3pt]
&\begin{aligned}
    s_{ij} ={}& - \frac{2\pi G}{mV} \sum_{\mathbf{k},\lambda} \sum_{\mathbf{k}',\lambda'} \alpha^{\lambda}(\mathbf{k}) \alpha^{\lambda'}(\mathbf{k}') e^{i(k+k') \cdot x} \\ 
    &\times \Big( \epsilon_{i}^{\lambda}(\mathbf{k}) \epsilon_{j}^{\lambda'}(\mathbf{k}') - \frac{1}{3} \delta_{ij} \epsilon_{k}^{\lambda}(\mathbf{k}) \epsilon_{k}^{\lambda'}(\mathbf{k}') \Big) + \mathrm{c.c.}\ .
\end{aligned}
\end{align}

In order to derive the timing residual two-point correlation function, we first calculate the two-point correlation functions of the metric perturbations.
The $\Psi$-$\Psi$ correlation function is
\begin{align}
    &\left< \Psi(x) \Psi(x') \right> = \frac{\pi^2 G^2 \bar{\rho}^2}{9m^4} \sum_{\mathbf{k},\lambda} \sum_{\mathbf{k}',\lambda'} P(\mathbf{k},\lambda) P(\mathbf{k}',\lambda') \\
    &\times \epsilon_{i}^{\lambda}(\mathbf{k}) \epsilon_{j}^{\lambda*}(\mathbf{k}) \epsilon_{i}^{\lambda'}(\mathbf{k}') \epsilon_{j}^{\lambda'*}(\mathbf{k}')  \cos \big( (k+k') \cdot (x-x') \big), \notag
\end{align}
where we have employed the four-point correlator Eq.~\eqref{4-point a adagger correlator}.
Assuming equipartition over the three polarizations, $P(\mathbf{k},\lambda) = P(\mathbf{k})P(\lambda) = P(\mathbf{k})/3$, the polarization sum gives $\sum_{\lambda} P(\lambda)\epsilon_{\mu}^{\lambda}(\mathbf{k})  \epsilon_{\nu}^{\lambda*}(\mathbf{k})  \simeq \eta_{\mu\nu}/3$, using the completeness relation, Eq.~\eqref{completeness relation}.
Replacing the discrete sums over momenta with integrals, the correlation function becomes
\begin{align}\
\begin{aligned}
    \left< \Psi(x) \Psi(x') \right> = \frac{\pi^2 G^2 \bar{\rho}^2}{27m^4} \int d^3 k \int d^3 k' f(\mathbf{k}) f(\mathbf{k}') \\
    \times \cos \big( (k+k') \cdot (x-x') \big).
\end{aligned}
\end{align}

To carry out the momentum space integration, we assume that the momentum of the local dark matter particles follows a Maxwell-Boltzmann distribution.
In the rest frame of the solar system, the distribution is centered at $\mathbf{k} = m\bar{\mathbf{v}}$ where $\bar{\mathbf{v}}$ is the mean velocity of the local dark matter wind due to the motion of the solar system around the galactic center:
\begin{align}\label{momentum PDF}
    f(\mathbf{k}) = \bigg(\frac{1}{2\pi m v_0^2}\bigg)^{3/2} \mathrm{exp}\bigg[ \frac{-|\mathbf{k}-m\bar{\mathbf{v}}|^2}{2mv_0^2} \bigg].
\end{align}
Expanding the phase of the cosine in the non-relativistic limit, the integration yields:
\begin{align}
\begin{aligned}
    \left< \Psi(x) \Psi(x') \right> = \frac{\pi^2 G^2 \bar{\rho}^2}{27m^4} e^{- |\Delta\mathbf{x}|^2 / \ell^2} \cos \big( 2m (\Delta t - \bar{\mathbf{v}} \cdot \Delta \mathbf{x}) \big),
\end{aligned}
\end{align}
where we have dropped all the $\mathcal{O}(\Delta t / \tau)$ terms as $\Delta t \ll \tau$ for pulsar timing observations of the fast mode.

Similarly, we calculate the $\Psi$-$s_{ij}$ and the $s_{ij}$-$s_{kl}$ correlation functions:
\begin{align}
    \left< \Psi(x) s_{ij}(x') \right> &= 0, \\
    \left< s_{ij}(x) s_{kl}(x') \right> &= \frac{8\pi^2 G^2 \bar{\rho}^2}{9m^4} 
    \big( \delta_{ik} \delta_{jl} + \delta_{il} \delta_{jk} - \frac{2}{3} \delta_{ij} \delta_{kl} \big) \notag \\
    \times\,& e^{- |\Delta\mathbf{x}|^2 / \ell^2} \cos \big( 2m (\Delta t - \bar{\mathbf{v}} \cdot \Delta \mathbf{x}) \big).
\end{align}
The cross $\Psi$-$s_{ij}$ correlation vanishes and $\Psi$-$\Psi$ and $s_{ij}$-$s_{kl}$ correlation functions share the same spacetime structure -- oscillatory in spacetime but with a suppression when the spatial separation exceeds the coherence length.
This is consistent with our discussion in Sec.~\ref{sec2} that correlation declines and the fields exhibit more stochasticity beyond the coherence scales.

With Eqs.~\eqref{fast mode redshift} and~\eqref{residual correlation function}, we obtain the fast mode timing residual two-point correlation function:~\footnote{The integration Eq.~\eqref{residual correlation function}, where the integrand is in the form of a cosine, actually yields four rather than one cosines:
\begin{align}
\begin{aligned}
    \int_0^t d\tau \int_0^{t'} d\tau' \cos (2m(\tau-\tau')+\varphi) = \cos (2m (t-t')+\varphi) \\
    - \cos (2mt +\varphi) 
    - \cos (-2mt' +\varphi) + \cos \varphi.
\end{aligned}
\end{align}
Here we only reserve the first cosine with stationary time dependence. 
The latter three will be absorbed by the timing model fitting of pulsar timing arrays and make no difference to the analysis. 
}
\begin{align}\label{fast mode timing residual correlation fucntion final result}
\begin{aligned}
    \left< r_a(t) r_b(t') \right> &= \mathcal{A}_{ab}\, \mathrm{Re} \Big[ U_{ab}\, e^{i2m(t-t')} \Big],
\end{aligned}
\end{align}
where $\mathcal{A}_{ab}$ and $U_{ab}$ are defined as
\begin{align}
    \mathcal{A}_{ab} &= - \frac{5\pi^2 G^2 \bar{\rho}^2}{36m^6} \bigg( 1 - \frac{16}{5} \big( \hat{\mathbf{n}}_a \cdot \hat{\mathbf{n}}_b \big)^2 \bigg), \label{A_ab} \\[5pt]
    U_{ab} &= 1 - e^{- d_b^2 / \ell^2} e^{i2m ( d_b + \bar{\mathbf{v}} \cdot d_b \hat{\mathbf{n}}_b )}
    - e^{- d_a^2 / \ell^2} e^{-i2m ( d_a + \bar{\mathbf{v}} \cdot d_a \hat{\mathbf{n}}_a )} \notag \\ 
    &+ e^{- |d_a\hat{\mathbf{n}}_a-d_b\hat{\mathbf{n}}_b|^2 / \ell^2} e^{i2m ( d_b - d_a + \bar{\mathbf{v}} \cdot ( d_b \hat{\mathbf{n}}_b - d_a \hat{\mathbf{n}}_a ) )}. \label{U_ab}
\end{align}
The magnitude of the fast mode timing residuals is $\mathcal{O}(G\bar{\rho}/m^3)$, indicated by $\mathcal{A}_{ab}$, consistent with our previous result, Eq.~\eqref{magnitude of timing residuals}.
The four terms in $U_{ab}$ correspond respectively to correlations between observer-observer, observer-pulsar $b$, the observer-pulsar $a$, and the pulsar $a$-pulsar $b$.

The time dependence of the correlation function is stationary and harmonic, at the fast mode frequency $\omega = 2m$, leading to a deterministic signal in time.
This is a consequence of the light travel time from the pulsars to the Earth always remaining well below the coherence time of dark matter in the $10^{-24}-10^{-22}\, \mathrm{eV}$ mass range, as we have discussed in Sec.~\ref{sec1}.

This monochromaticity can be used to simplify the pulsar timing analysis of the fast mode.
In practice, where each pulsar's timing residuals are measured at discrete observation times within a finite time span, the covariance matrix is defined by sampling the two-point correlation function, Eq.~\eqref{fast mode timing residual correlation fucntion final result}, onto these discrete time series.
Supposing there are $N_\mathrm{P}$ pulsars involved, the covariance matrix has at most $2N_\mathrm{P}$ nonzero eigenvalues.
Each eigenvector that belongs to a nonzero eigenvalue is the sampling of a set of trigonometric functions labeled by the $N_\mathrm{P}$ pulsars at angular frequency $2m$.
This feature corresponds to the fact that the timing residuals from each pulsar must follow a harmonic time dependence regardless of the values taken by the random variables $\{\alpha^\lambda(\mathbf{k})\}$.
Therefore, the computation time can be significantly reduced by projecting the observed data onto these eigenvectors with nonzero eigenvalues, since the other components in the observed data do not contain information about vector dark matter at all.
We examine the eigenvalue problem of this monochromatic covariance matrix in App.~\ref{apdx4}.

\begin{figure}
    \centering
    \includegraphics[width=\linewidth]{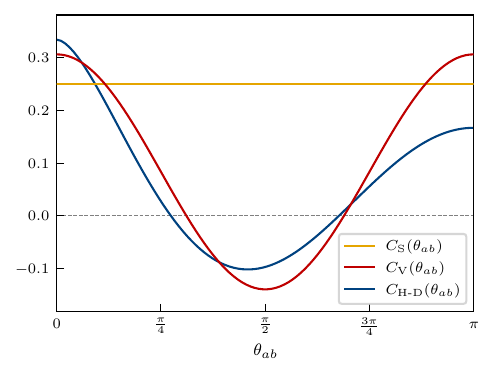}
    \caption{Dependence on $\theta_{ab}$ of $\mathcal{A}_{ab}$ for scalar dark matter, $C_{\mathrm{S}}(\theta_{ab})$, and vector dark matter, $C_{\mathrm{V}}(\theta_{ab})$, and the Hellings-Downs curve for a stochastic gravitational wave background, $C_{\mathrm{H\text{-}D}}(\theta_{ab})$.
    $C_{\mathrm{S}}(\theta_{ab})$ and $C_{\mathrm{V}}(\theta_{ab})$ are normalized as the factor apart from the $\pi^2 G^2 \bar{\rho}^2/m^6$ in $\mathcal{A}_{ab}$ such that $C_{\mathrm{S}}(\theta_{ab}) = 1/4$ and $C_{\mathrm{V}}(\theta_{ab}) = - (5 - 16\cos^2\theta_{ab})/36$.}
    \label{angular dependence}
\end{figure}

The non-trivial dependence of $\mathcal{A}_{ab}$ on the angle between pulsars encodes the key distinguishing feature of vector dark matter. 
This particular form of $\mathcal{A}_{ab}$ is a consequence of a combination of the scalar, vector, and tensor perturbation contributions in the metric.
In contrast, the fast mode timing residual of scalar dark matter only carries one scalar perturbation in the leading order, resulting in a constant $\mathcal{A}_{ab}$~\cite{Dror:2025_PTA}.
For a stochastic gravitational wave background, its tensor perturbations follow different statistics, which lead to angular dependence known as the Hellings-Downs curve~\cite{Hellings:1983fr}.
We plot the dependence on the angle between two pulsars, $\theta_{ab} \equiv \mathrm{arccos}(\hat{\mathbf{n}}_a \cdot \hat{\mathbf{n}}_b)$, of $\mathcal{A}_{ab}$ for  scalar and vector dark matter, and compare with the Hellings-Downs curve for a stochastic gravitational wave background in Fig.~\ref{angular dependence}.
Apart from $\mathcal{A}_{ab}$, the fast mode timing residual two-point correlation functions for scalar and vector dark matter are exactly the same~\cite{Dror:2025_PTA}.

The pulsar locations also control the loss of correlation through exponential decay factors in $U_{ab}$, which are a manifestation of the field of mass $m \sim 10^{-24} - 10^{-22}\, \mathrm{eV}$ being spatially stochastic on pulsar timing arrays.
In the $\ell \rightarrow 0$ limit, metric perturbations at different locations are completely uncorrelated, and only self-correlations survive in the timing residual correlation function:
\begin{align}
\begin{aligned}
    \left< r_a(t) r_b(t') \right> = \mathcal{A}_{ab} \big( 1 + \delta_{ab} \big) \cos \big( 2m(t-t') \big).
\end{aligned}
\end{align}
This limit corresponds to the fully uncorrelated regime.
In the opposite limit $\ell \rightarrow \infty$, metric perturbations over the spacetime are fully correlated. 
This reduces the timing residual correlation function to:
\begin{align}\label{fully correlated}
    &\left< r_a(t) r_b(t') \right> \\
    &= 4\mathcal{A}_{ab} \sin \big( m \big( d_b + \bar{\mathbf{v}} \cdot d_b \hat{\mathbf{n}}_b \big) \big) \sin \big( m \big( d_a + \bar{\mathbf{v}} \cdot d_a \hat{\mathbf{n}}_a \big) \big) \notag \\ 
    &\times \cos \big( 2m \big( t-t' \big) +m \big( d_b - d_a + \bar{\mathbf{v}} \cdot \big( d_b \hat{\mathbf{n}}_b - d_a \hat{\mathbf{n}}_a \big) \big) \big). \notag
\end{align}
This limit corresponds to the single-plane-wave formulation, Eq.~\eqref{coherent classical field}, with $\beta_i$ generated by the Gaussian distribution, Eq.~\eqref{beta_i}, and $\mathbf{p}$ fixed at $m\bar{\mathbf{v}}$.
If also $d_a, d_b \ll 1/m$, the sines in Eq.~\eqref{fully correlated} give zero and the correlation function vanishes.
In this case, the metric remains constant over the propagation of photons from the pulsars to the Earth, leaving us no signal at all.

The pulsar locations in the phases in $U_{ab}$ affect both the amplitude and the phase of the harmonic oscillation of the timing residual correlation function with time.
In particular, $U_{ab}$ is anisotropic as the phases depend on the angles between the pulsar orientations and the mean velocity of the local dark matter wind $\bar{\mathbf{v}}$.
This anisotropy disappears when pulsars are well within $1/m|\bar{\mathbf{v}}|$, a distance comparable to the coherence length $\ell$, around the observer. 
One may include the pulsar locations in the analysis by marginalizing over their values informed by a Gaussian prior with a mean set by the best fit value and standard deviation set by its uncertainty.  
However, resolving the influence of the phases requires measurements of pulsar distances to a precision at least better than $1/m = 6.4 \times 10^{-1}\, \mathrm{pc}\, (m/10^{-23}\, \mathrm{eV})^{-1}$, which is far below the distance uncertainties for almost all pulsars of our interest~\cite{NANOGrav:2023_datarelease}. 
In practice, one may simply assign a uniformly distributed random phase $\phi$ to each pulsar, i.e., rewriting $U_{ab}$ as:
\begin{align}
\begin{aligned}
    U_{ab} ={}& 1 - e^{- d_b^2 / \ell^2}\, e^{i\phi_b} - e^{- d_a^2 / \ell^2}\, e^{i\phi_a} \\
    &+ e^{- |d_a\hat{\mathbf{n}}_a-d_b\hat{\mathbf{n}}_b|^2 / \ell^2}\, e^{i(\phi_a+\phi_b)},
    \end{aligned}
\end{align}
and marginalize over them in a likelihood-based analysis. These large distance uncertainties wipe out the anisotropy of $U_{ab}$, eliminating observability of the mean local dark matter wind velocity in pulsar timing arrays. 

Our derivation of Eqs.~\eqref{fast mode timing residual correlation fucntion final result},~\eqref{A_ab}, and~\eqref{U_ab} assumes the spatial homogeneity of the properties of vector dark matter, neglecting the deformation of the wavefunctions of dark matter particles due to the gravitational potential in the halo.
For the Navarro–Frenk–White profile of the Milky Way, within 1 kpc, $v_0$ varies about 2\%, which can be safely neglected, but $\bar{\rho}$ varies about 20\%, which introduces a considerable correction.
The spatial variation of the dark matter density can be incorporated in our formalism by replacing $\bar{\rho}^2$ in $\mathcal{A}_{ab}$ with $\bar{\rho}(\mathbf{0})^2$, $\bar{\rho}(\mathbf{0})\bar{\rho}(d_b\hat{\mathbf{n}}_b)$, $\bar{\rho}(d_a\hat{\mathbf{n}}_a)\bar{\rho}(\mathbf{0})$, and $\bar{\rho}(d_a\hat{\mathbf{n}}_a)\bar{\rho}(d_b\hat{\mathbf{n}}_b)$ for each of the four terms in $U_{ab}$.~\footnote{A more precise approach would take into account gravitational clustering by swapping the free wavefunctions making up of Eq.~\eqref{field operator} for solutions of the Schr\"{o}dinger-Poission equations~\cite{Kim:2021_DMclustering} and introducing a spatial dependence to the momentum distribution $P(\mathbf{k})$, which complicates the analysis.}

\section{\label{sec5}The slow mode pulsar timing signal}

We now study the timing residual two-point correlation function for the slow mode.
We work in the {\em transverse gauge}, where only the scalar potential $\Phi$ contributes at leading order in the velocity expansion (see Tab.~\ref{parametric sizes}).
Returning to the redshift expression in Eq.~\eqref{redshift}, we retain only the contribution from $\Phi$ in $R_{i0j0}$ and only the leading term in the velocity expansion of the line-of-sight integral for the slow mode, $(\partial_0 - \hat{n}_i \partial_i)^{-1} \simeq - (\hat{n}_i \partial_i)^{-1}$.
Altogether, this yields the redshift expression for the slow mode:
\begin{align}\label{slow mode redshift}
\begin{aligned}
    z_a(t) = \hat{\mathbf{n}}_a \cdot  \int^t dt'\, \Big( \nabla \Phi(t',\mathbf{0}) -  \nabla \Phi(t'-d_a,d_a\hat{\mathbf{n}}_a) \Big).
\end{aligned}
\end{align}
Using the structure of the slow mode of $T_{\mu\nu}$ in Tab.~\ref{parametric sizes} and the scaling of Eq.~\eqref{sizes of differential operators}, the solution for $\Phi$ from the Einstein's equations in the transverse gauge provided in App.~\ref{apdx1} can be simplified, to the leading order in the velocity expansion:
\begin{align}\label{simp Phi}
    \Phi = 4\pi G \frac{1}{\nabla^2} T_{00}.
\end{align}

The slow mode term of $T_{00}$ to the leading order is:
\begin{align}
    &\quad T_{00} = \frac{1}{2} \Big( |\dot{\mathbf{A}}|^2 + m^2 |\mathbf{A}|^2 \Big) \\
    &= \frac{m}{2V} \sum_{\mathbf{k},\lambda} \sum_{\mathbf{k}',\lambda'} \alpha^{\lambda}(\mathbf{k}) \alpha^{\lambda'*}(\mathbf{k}') \epsilon_{i}^{\lambda}(\mathbf{k}) \epsilon_i^{\lambda'*}(\mathbf{k}') e^{i(k-k') \cdot x} + \mathrm{c.c.} \notag
\end{align}
We then calculate the two-point correlation function of the density fluctuation, $\delta T_{00} \equiv T_{00} - \left< T_{00} \right>$, to which pulsar timing arrays are primarily sensitive.
Assuming equipartition over the three polarizations and using the four-point correlator of Eq.~\eqref{4-point a adagger correlator}, we find:
\begin{align}\label{density correlation function}
\begin{aligned}
    \left< \delta T_{00}(x) \delta T_{00}(x') \right> = \frac{\bar{\rho}^2}{3} \sum_{\mathbf{k},\mathbf{k}'} P(\mathbf{k}) P(\mathbf{k}') \cos \big( (k-k') \cdot \Delta x \big),
\end{aligned}
\end{align}
where $\Delta x \equiv x-x'$.
Transforming this correlation function into the momentum space gives:
\begin{align}\label{density correlation function in k}
    &\big< \widetilde{\delta T}_{00}(k) \widetilde{\delta T}^*_{00}(k') \big> = (2\pi)^4 \delta^4 (k-k') S_{\delta \rho}(k),
\end{align}
where $S_{\delta \rho}(k)$ is the power spectrum of the density fluctuations:
\begin{align}\label{density fluctuation power spectrum}
    S_{\delta \rho}(k) = \frac{\bar{\rho}^2}{3} \sum_{\mathbf{k'},\mathbf{k}''} \hspace{-0.1cm}P(\mathbf{k}') P(\mathbf{k}'') (2\pi)^4 \delta^4 (k - k' + k'').
\end{align}

As argued in Sec.~\ref{sec3}, the slow mode timing residual two-point correlation function of vector dark matter resembles that of scalar dark matter.
In particular, since the slow modes of scalar and vector dark matter share the same parametric structures (see Tab.~\ref{parametric sizes}), their slow mode timing residuals, to the leading order in the velocity expansion, both arise from density fluctuations $\delta T_{00}$ through Eqs.~\eqref{slow mode redshift} and~\eqref{simp Phi} in the transverse gauge.
Comparing Eqs.~\eqref{density correlation function in k} and~\eqref{density fluctuation power spectrum} with the power spectrum of the slow mode density fluctuations of scalar dark matter~\cite{Kim:2023_PTA}, we find that they are identical except for an overall factor of 3 in amplitude. 
The slow mode timing residual two-point correlation function of vector dark matter is therefore identical to that of scalar dark matter, but with an amplitude smaller by a factor of 3. 

Accordingly, the slow mode analysis in Ref.~\cite{Kim:2023_PTA} can be directly applied to vector dark matter, with constraints on the local dark matter density weaker by $1/\sqrt{3}$ (the current bounds are of $ {\cal O} ( 10 ^3 ) $ times the expected local density).

This difference can be understood intuitively since vector dark matter has three polarizations, each of which exhibits its own slow mode. 
These slow modes are uncorrelated and hence do not generically constructively interfere in $T_{00}$. 
The same result would apply if dark matter consisted of three distinct scalar fields with identical masses and velocity distributions.

To understand the coherence properties of the slow mode, we carry out the momentum integral in Eq.~\eqref{density correlation function} with the Maxwell-Boltzmann momentum distribution, taking the non-relativistic limit $\omega_{\mathbf{k}} \simeq m + |\mathbf{k}|^2/2m$:
\begin{align}
\begin{aligned}
    &\left< \delta T_{00}(x) \delta T_{00}(x') \right>  \\
    &= \frac{\bar{\rho} ^2 / 3 }{(1+{\Delta t}^2/\tau^2)^{3/2}} \, \mathrm{exp} \bigg( -\frac{\big| \Delta \mathbf{x} - \bar{\mathbf{v}} \Delta t \big|^2 / \ell^2}{ 1+{\Delta t}^2/\tau^2  } \bigg).
\end{aligned}
\end{align}
Unlike the fast mode, the slow mode shows no oscillatory spacetime dependence. 
This is a consequence of the slow mode arising entirely due to the stochasticity of the vector field. 
Nonetheless, the density fluctuation two-point function is still strongly suppressed when $\Delta t \gg \tau$ and $|\Delta \mathbf{x}| \gg \ell$. 
For the mass range relevant for the slow mode ($ 10^{-18}-10^{-16}\, \mathrm{eV}$), the Earth-pulsar separations are much larger than the coherence scales both spatially and temporally. 
Hence, the density fluctuations at the Earth and at different pulsars are effectively uncorrelated, and only self-correlations contribute to the slow mode timing residual correlation function.

\section{\label{sec6}Conclusion}
In this work, we presented a statistical treatment of the influence of ultralight vector dark matter on pulsar timing arrays. Beginning with a quantum mechanical description of the field, we showed that it reduces to a stochastic classical field with well-defined statistical properties. This allowed us to treat ultralight vector dark matter as a stochastic background --- rather than a deterministic signal --- and compute the corresponding statistics of its induced metric perturbations. From these results, we derived our key observable: the two-point correlation function of pulsar timing residuals, for both the fast and the slow modes. Our results are structured to enable future searches for ultralight vector dark matter.

The fast mode, which can be uncovered with pulsar timing arrays if dark matter has mass in the range $10^{-24}-10^{-22}\, \mathrm{eV}$, encodes the most distinctive observational signature of vector dark matter: a unique angular dependence in the timing residual two-point correlation function, distinct from both scalar dark matter and stochastic gravitational wave backgrounds. Fast mode signals are coherent in time but stochastic in space, resulting in an exponential suppression of correlations between the Earth and pulsars (and between pulsars) when their spatial separations exceed the field's coherence length.

The slow mode, which can be uncovered with pulsar timing arrays if dark matter mass lies in the $ 10^{-18}-10^{-16}\, \mathrm{eV}$ range, resembles the scalar dark matter signal but with a reduced amplitude. Accordingly, ongoing and future searches for the slow mode of scalar dark matter can simultaneously test for the vector dark matter. 

Drawing on our observability analysis and the sensitivities that current observations have achieved for scalar dark matter, we expect the pulsar timing searches for vector dark matter to achieve a sensitivity around the expected local dark matter density ($\simeq 0.3\, \mathrm{GeV/cm^3}$) in the $10^{-24}-10^{-22}\, \mathrm{eV}$ mass range by probing the fast mode signals, and about $\mathcal{O}(10^3)$ times the expected density in the $10^{-18}-10^{-16}\, \mathrm{eV}$ range through the slow mode signals.

We now briefly remark on some potential directions for future work. 
First, we emphasize once more that our results assumed the vector polarizations are equally populated, i.e., equipartition. 
While this is a reasonable expectation given the nonlinear gravitational collapse into galactic halos and their subsequent virialization, it is conceivable that any initial polarization preference may survive to the present day. 
This would influence the timing residual two-point correlation functions, and, in principle, could be detected by pulsar timing arrays.

Second, we note that galactic dark matter halos composed of ultralight vector spin-$1$ particles contain sizable tensor perturbations and hence act as gravitational wave sources. 
However, efficient gravitational wave production requires matching the source's frequency and wavenumber, since a mismatch leads to destructive interference among the light-speed propagating tensor perturbations generated from different regions of the source.
For the fast mode, $\omega \simeq 2m$ is about $10^3$ times larger than the typical wavenumber $|\mathbf{k}| \sim mv_0$. 
For the slow mode, $\omega \lesssim mv_0^2$ is about $10^3$ times smaller than the typical wavenumber $|\mathbf{k}| \sim mv_0$. 
Thus, both modes are highly inefficient at producing gravitational waves. Nonetheless, there may still be opportunities to probe these propagating tensor modes.

Finally, the framework developed in this work is applicable to other avenues of gravitational direct detection of ultralight vector dark matter, particularly at scales comparable to the coherence scales of dark matter.
It could also be extended to study the gravitational effects of spin-$2$ (or even higher spin) ultralight dark matter. 
It would be useful to explore how such models may be observationally distinguished from scalar or vector dark matter. 

\begin{acknowledgements}
The authors thank Fengwei Yang, Kimberly Boddy, and Nicholas Rodd for useful discussions. The research of JD is supported in part by the U.S. Department of Energy grant number DE-SC0025569. The research of QW is supported in part by the Institute of High Energy Physics and Astrophysics 2024 summer research fellowship.
\end{acknowledgements}

\bibliography{paper}

\clearpage
\onecolumngrid
\appendix

\section{\label{apdx1}Solutions to Einstein's Equations in Different Gauges}

In this appendix, we provide the solutions to Einstein's equations in the weak-field limit in both the transverse and the synchronous gauge.
We further reduce the solutions, retaining only the leading-order contributions for the fast and the slow modes, respectively, based on the structure of the energy-momentum tensor, the metric perturbations, and the timing residuals as shown in Tab.~\ref{parametric sizes}.

In the transverse gauge ($\partial_i w_i = \partial_i s_{ij} = 0$), the solutions are
\begin{align}
    \Psi &= 4\pi G \frac{1}{\nabla^2} T_{00}, \\
    \Phi &= 4\pi G \frac{1}{\nabla^2} \bigg( T_{kk} +  \bigg( 1- 3 \frac{\partial_0^2}{\nabla^2} \bigg) T_{00} \bigg), \label{Phi trans} \\
    w_{i} &= - 16\pi G\frac{1}{\nabla^2} P_{ij}T_{0j}, \\ 
    s_{ij} &= - 8\pi G \frac{1}{\Box} \Lambda_{ij,kl} T_{kl},
\end{align}
where $P_{ij}$ is the transverse projector in the position space:
\begin{align}\label{Pij}
    P_{ij} = \delta_{ij} - \frac{\partial_i \partial_j}{\nabla^2},
\end{align}
and $\Lambda_{ij,kl}$ is the transverse-traceless projector in the position space:
\begin{align}\label{Lambdaijkl}
    \Lambda_{ij,kl} = P_{ik} P_{jl} - \frac{1}{2} P_{ij} P_{kl}.
\end{align}

In the synchronous gauge ($\Phi = w_i = 0$), the solutions are
\begin{align}
    \Psi &= \frac{4}{3} \pi G \frac{1}{\partial_0^2} (T_{00} + T_{kk}), \label{Psi sync} \\
    s_{ij} &= - 8\pi G \frac{1}{\Box} \bigg( \pi_{ij} + \frac{2}{3} \delta_{ij} T_{00} - \frac{\partial_i}{\partial_0} T_{0j} - \frac{\partial_j}{\partial_0} T_{0i} + \frac{1}{2} \frac{\nabla^2}{\partial_0^2} \bigg( \frac{\partial_i \partial_j}{\nabla^2} - \frac{1}{3} \delta_{ij} \bigg) (T_{00} + T_{kk})  \bigg). \label{sij sync}
\end{align}

Note that in the derivation of the above solutions, we have used energy-momentum conservation:
\begin{align}
    \partial^\mu T_{\mu\nu} = 0 \quad \Longrightarrow \quad \partial_0 T_{0\nu} = \partial_i T_{i\nu} \quad \Longrightarrow \quad \partial_0^2 T_{00} = \partial_0 \partial_i T_{i0} = \partial_i \partial_j T_{ij},
\end{align}
to simplify the expressions.

According to the discussion in Sec.~\ref{sec2}, we calculate the fast mode timing residuals in the synchronous gauge, where the leading-order contributions are from both $\Psi$ and $s_{ij}$.
For the fast mode, to the right of $1/\partial_0^2$ in Eq.~\eqref{Psi sync}, $T_{kk} \sim \mathcal{O}(\bar{\rho})$ is larger than $T_{00} \sim \mathcal{O}(\bar{\rho}v_0^2)$, and is thus the leading-order contribution to $\Psi$. 
In Eq.~\eqref{sij sync}, to the right of the $1/\Box$ operator, $\pi_{ij} \sim \mathcal{O}(\bar{\rho})$, $T_{00} \sim \mathcal{O}(\bar{\rho}v_0^2)$, and all other terms are suppressed with powers of $v_0$ introduced by spatial derivatives.
The leading-order contribution to $s_{ij}$ is thus from $\pi_{ij}$.
Accordingly, the solutions for the fast mode perturbations $\Psi$ and $s_{ij}$ in the synchronous gauge are reduced to:
\begin{align}
    \Psi &= \frac{4}{3} \pi G \frac{1}{\partial_0^2} T_{kk}, \\
    s_{ij} &= - 8\pi G \frac{1}{\Box} \pi_{ij}.
\end{align}

For the slow mode, the timing residuals are studied in the transverse gauge, where the leading-order contribution is only from $\Phi$.
In Eq.~\eqref{Phi trans}, to the right of the $1/\nabla^2$ operator, $T_{kk} \sim \mathcal{O}(\bar{\rho}v_0^2)$, $T_{00} \sim \mathcal{O}(\bar{\rho})$, and $(\partial_0^2/\nabla^2)T_{00} \sim \mathcal{O}(\bar{\rho}v_0^2)$.
Hence, $T_{00}$ is the leading-order term in Eq.~\eqref{Phi trans}, and the solution is reduced to:
\begin{align}
    \Phi &= 4\pi G \frac{1}{\nabla^2} T_{00}.
\end{align}

\section{\label{apdx2}Impact of Metric Perturbations on Pulsar Timings}

Metric perturbations affect pulsar timings by redshifting the pulsar emissions received by observers, which accumulate to timing residuals over time.
In this appendix, we discuss the effects of metric perturbations on the frequencies of pulsar emissions and provide a derivation for Eq.~\eqref{redshift}.

Suppose we are in a coordinate system where the metric satisfies the weak-field limit, and the motion of both the pulsar and the observer is non-relativistic.
We focus solely on the impact of gravity such that the pulsar, the observer, and the emitted photon all move along geodesics.
For a photon that propagates along the geodesic $x^{\mu}(\lambda)$ from the pulsar $x^\mu_{\mathrm{e}}=x^{\mu}(\lambda_{\mathrm{e}})$ to the observer $x^\mu_{\mathrm{o}}=x^{\mu}(\lambda_{\mathrm{o}})$, where the affine parameter $\lambda$ is chosen such that the momentum of the photon is $p^\mu(\lambda)=dx^{\mu}(\lambda)/d\lambda$, its emission frequency, $\omega_\mathrm{e}$, and the frequency observed by the observer, $\omega_\mathrm{o}$, are
\begin{align}
    \omega_\mathrm{e} &= - g_{\mu\nu}(x_\mathrm{e}) p^\mu(\lambda_\mathrm{e}) U^\nu_\mathrm{P}(t_\mathrm{e}) \\
    \omega_\mathrm{o} &= - g_{\mu\nu}(x_\mathrm{o}) p^\mu(\lambda_\mathrm{o}) U^\nu_\mathrm{O}(t_\mathrm{o}),
\end{align}
where $U^\nu_\mathrm{P}$ and $U^\nu_\mathrm{O}$ are the 4-velocity of the pulsar and the observer, evaluated at the time of emission, $t_\mathrm{e}$, and the time of arrival, $t_\mathrm{o}$.
Thus, the frequency change is
\begin{align}\label{frequency change}
\begin{aligned}
    \Delta \omega ={}& \omega_\mathrm{o} - \omega_\mathrm{e} \\
    ={}& - \big( \eta_{\mu\nu} + h_{\mu\nu}(x_\mathrm{o}) \big) \Big( p^{(0)\mu}(\lambda_\mathrm{o}) + p^{(1)\mu}(\lambda_\mathrm{o}) \Big) \Big( U^{(0)\nu}_\mathrm{O}(t_\mathrm{o}) + U^{(1)\nu}_\mathrm{O}(t_\mathrm{o}) \Big) \\
    &+ \big( \eta_{\mu\nu} + h_{\mu\nu}(x_\mathrm{e}) \big) \Big( p^{(0)\mu}(\lambda_\mathrm{e}) + p^{(1)\mu}(\lambda_\mathrm{e}) \Big) \Big( U^{(0)\nu}_\mathrm{P}(t_\mathrm{e}) + U^{(1)\nu}_\mathrm{P}(t_\mathrm{e}) \Big) \\
    ={}& \Delta \omega^{(0)} - \eta_{\mu\nu} p^{(0)\mu} \Big( U^{(1)\nu}_\mathrm{O}(t_\mathrm{o}) - U^{(1)\nu}_\mathrm{P}(t_\mathrm{e}) \Big) - \eta_{\mu\nu} \Big( p^{(1)\mu}(\lambda_\mathrm{o}) U^{(0)\nu}_\mathrm{O}(t_\mathrm{o}) - p^{(1)\mu}(\lambda_\mathrm{e}) U^{(0)\nu}_\mathrm{P}(t_\mathrm{e}) \Big) \\
    &- p^{(0)\mu} \Big(h_{\mu\nu}(x_\mathrm{o}) U^{(0)\nu}_\mathrm{O}(t_\mathrm{o}) - h_{\mu\nu}(x_\mathrm{e}) U^{(0)\nu}_\mathrm{P}(t_\mathrm{e}) \Big),
\end{aligned}
\end{align}
where we expand each quantity in terms of $h_{\mu\nu}$ to the first order in the first equality and keep $\Delta\omega$ to the first order of $h_{\mu\nu}$ in the second equality, with the superscript (0), (1) denoting respectively the unperturbed quantities and the first-order corrections.
The zeroth-order frequency change $\Delta \omega^{(0)} = - \eta_{\mu\nu} p^{(0)\mu} \Big( U^{(0)\mu}_\mathrm{O}(t_\mathrm{o}) -U^{(0)\mu}_\mathrm{P}(t_\mathrm{e}) \Big)$ is the ordinary Doppler effect arising from the unperturbed relative motion between the pulsar and the observer.
We have dropped the argument $\lambda$ for the unperturbed photon momentum $p^{(0)\mu}$ since it is constant along the photon geodesic.

The first-order frequency change $\Delta \omega^{(1)}$, according to the last expression in Eq.~\eqref{frequency change}, is a sum of three different effects:
\begin{enumerate}
    \item the Doppler effect between the observer and the pulsar due to gravitational accelerations,
    \item the perturbations to the photon geodesic and momentum during propagation (the generalized integrated Sachs-Wolfe effect), and
    \item the non-trivial metrics at the pulsar and the observer.
\end{enumerate}
They respectively correspond to keeping only the first-order corrections of $U^{\mu}_{\{\mathrm{P},\mathrm{O}\}}$, $p^{\mu}(\lambda)$ and $g_{\mu\nu}$ in the expression of $\Delta\omega$.

The expression of the redshift in terms of metric perturbations can certainly be obtained following the decomposition in Eq.~\eqref{frequency change}.
However, we provide a neater derivation which leads directly to the compact expression of Eq.~\eqref{redshift} by examining the evolution of the observed photon frequency along its path.

Consider a one-parameter family of non-relativistic free-falling observers labeled by $\lambda$, where the worldline of observer $\lambda$ intersects with the photon geodesic at $x^{\mu}(\lambda)$.
The pulsar and the observer on the Earth are thus identified as observer $\lambda_\mathrm{e}$ and observer $\lambda_\mathrm{o}$.
In the vicinity of the photon geodesic, the collection of the worldlines of these observers defines a two-dimensional surface, on which their 4-velocities form a vector field, $U^\mu(t,\lambda)$, parametrized by time $t$ and the label of the observers $\lambda$ and representing the 4-velocity of observer $\lambda$ at coordinate time $t$.

Define $\omega(\lambda)$ as the frequency of the photon measured by observer $\lambda$ at $x^{\mu}(\lambda)$:
\begin{align}
    \omega(\lambda) = - g_{\mu\nu}(x(\lambda))\, p^\mu(\lambda)\, U^\nu(x^0(\lambda),\lambda).
\end{align}
We calculate its derivative $d\omega(\lambda)/d\lambda$ and retain terms up to the first order of $h_{\mu\nu}$:
\begin{align}\label{evolution of frequency}
\begin{aligned}
    \frac{d}{d\lambda} \omega(\lambda) ={}& \frac{D}{d\lambda} \omega(\lambda) \\
    ={}& - g_{\mu\nu}(x(\lambda))\, p^\mu(\lambda)\, \frac{D}{d\lambda} U^\nu(x^0(\lambda),\lambda) \\
    ={}& \frac{d}{d\lambda} \omega^{(0)}(\lambda) - \eta_{\mu\nu} p^{(0)\mu} \bigg( p^{(0)\rho}\, \Gamma_{\rho\sigma}^{\nu}(x(\lambda))\, U^{(0)\sigma}(x^0(\lambda),\lambda) + \frac{d}{d\lambda} U^{(1)\nu}(x^0(\lambda),\lambda) \bigg) \\
    &- \Big( \eta_{\mu\nu} p^{(1)\mu}(\lambda) + h_{\mu\nu}(x(\lambda))\, p^{(0)\mu} \Big) \frac{d}{d\lambda} U^{(0)\nu}(x^0(\lambda),\lambda) \\
    ={}& \frac{d}{d\lambda} \omega^{(0)}(\lambda) - \eta_{\mu\nu} p^{(0)\mu}  p^{(0)\rho} \Big( \Gamma_{\rho 0}^{\nu}(x(\lambda)) + \partial_\rho U^{(1)\nu}(x^0(\lambda),\lambda) \Big). 
\end{aligned}
\end{align}
In the first equality, we replace the derivative with the covariant derivative since $\omega(\lambda)$ is a scalar.
In the second equality, we take advantage of the metric compatibility of the connection, $Dg_{\mu\nu}/d\lambda = 0$, and the geodesic equation of the photon, $Dp^{\mu}(\lambda)/d\lambda = 0$, such that the only nonzero term is from the covariant derivative acting on the 4-velocity of the observers.
In the third equality, we expand the covariant derivative, drop the higher-order terms of $h_{\mu\nu}$, and collect the zeroth-order terms into $d\omega^{(0)}(\lambda)/d\lambda = - \eta_{\mu\nu} p^{(0)\mu} d U^{(0)\nu}(x^0(\lambda),\lambda)/d\lambda$.
In the last equality, we take $U^{(0)\nu}(x^0(\lambda),\lambda) = (1,0,0,0)$ in the first-order terms, neglecting the motion of the observers.
This is because the non-relativistic motion of the observers contributes only $\mathcal{O}(vh_{\mu\nu})$ to the first-order terms, with their typical velocity $v \ll 1$.

The first-order correction to the 4-velocity of any observer, $U^{(1)\nu}$, can be derived from its geodesic equation:
\begin{align}
\begin{aligned}
    0 &= \frac{D}{d\tau} U^{\nu} = \frac{d}{d\tau} U^{\nu} + \Gamma^\nu_{\rho\sigma} U^\rho U^\sigma = \frac{d}{d\tau} U^{(0)\nu} + \frac{d}{d\tau} U^{(1)\nu} + \Gamma^\nu_{\rho\sigma} U^{(0)\rho} U^{(0)\sigma}
\end{aligned}
\end{align}
where we also retain terms up to the first order.
The first order of this geodesic equation yields:
\begin{align}
    0 = \frac{d}{d\tau} U^{(1)\nu} + \Gamma^\nu_{\rho\sigma} U^{(0)\rho} U^{(0)\sigma} \quad\Longrightarrow\quad
    \frac{d}{dt} U^{(1)\nu} = \frac{d\tau}{dt} \frac{d}{d\tau} U^{(1)\nu} = \frac{1}{U^{(0)0}} \frac{d}{d\tau} U^{(1)\nu} = - \Gamma^\nu_{00},
\end{align}
where we have kept up to the first order and plugged in $U^{(0)\nu}(x^0(\lambda),\lambda) = (1,0,0,0)$ for the same reason as we have discussed above.
Then $U^{(1)\nu}$ can be obtained by performing a time integration on $- \Gamma^\nu_{00}$ along the unperturbed worldline of the observer, $\mathcal{C}_\mathrm{O}^{(0)}$, which, neglecting the motion of the observer, is equivalent to an integration simply with respect to time:
\begin{align}
    U^{(1)\nu} = -\int_{\mathcal{C}_\mathrm{O}^{(0)}} dt\,\Gamma^\nu_{00} = -\frac{1}{\partial_0}\Gamma^\nu_{00}.
\end{align}

Supposing the unperturbed propagation of the photon is in the $-\hat{\mathbf{n}}$ direction and again neglecting the motion of the pulsar, the unperturbed momentum of the photon is $p^{(0)\mu} = \omega_\mathrm{e} (1,-\hat{\mathbf{n}})$, with the emission frequency $\omega_\mathrm{e}$.
Plugging it and the expressions of $U^{(1)\nu}$ into the last expression in Eq.~\eqref{evolution of frequency}, we obtain the first-order term of the frequency evolution, $d\omega^{(1)}(\lambda)/d\lambda$:
\begin{align}
\begin{aligned}
    \frac{d}{d\lambda} \omega^{(1)}(\lambda) &= - \eta_{\mu\nu} p^{(0)\mu}  p^{(0)\rho} \bigg( \Gamma_{\rho 0}^{\nu}(x(\lambda)) - \frac{\partial_\rho}{\partial_0} \Gamma_{00}^{\nu}(x(\lambda)) \bigg) \\
    &= p^{(0)\mu}  p^{(0)\nu} \frac{1}{\partial_0} R_{\mu 0 \nu 0}(x(\lambda)) \\
    &= \omega_\mathrm{e}^2 \hat{n}_i \hat{n}_j \frac{1}{\partial_0} R_{i0j0}(x(\lambda)) ,
\end{aligned}
\end{align}
where $R_{\mu\nu\rho\sigma}$ is the Riemann curvature tensor and $R_{i0j0}$ are the only non-vanishing components of $R_{\mu 0 \nu 0}$ to the first order of $h_{\mu\nu}$.
Therefore, the redshift observed by the observer on the Earth is derived by performing an integral along the photon geodesic from the pulsar to the Earth:
\begin{align}\label{redshift derivation}
\begin{aligned}
    z &= \frac{\Delta\omega}{\omega_\mathrm{e}} \\
    &= z^{(0)} + \omega_\mathrm{e} \hat{n}_i \hat{n}_j \int_{\lambda_\mathrm{e}}^{\lambda_{\mathrm{o}}} d\lambda\, \frac{1}{\partial_0} R_{i0j0}(x(\lambda)) \\
    &= z^{(0)} + \hat{n}_i \hat{n}_j \int_{t_\mathrm{e}}^{t_{\mathrm{o}}} dt\, \frac{1}{\partial_0} R_{i0j0}(t,\mathbf{x}_\gamma^{(0)}(t))
\end{aligned}
\end{align}
where $z^{(0)}$ is the ordinary Doppler redshift, $\mathbf{x}_\gamma^{(0)}(t)$ is the zeroth-order trajectory of the emission, and in the last equality we change the integration variable from $\lambda$ to $t$, noticing that along the unperturbed photon geodesic: 
\begin{align}
    \frac{d}{d\lambda} x^{(0)0}(\lambda) = p^{(0)0} = \omega_\mathrm{e} \quad\Longrightarrow\quad dt = \omega_\mathrm{e} d\lambda \, .
\end{align}
The last expression of Eq.~\eqref{redshift derivation} gives Eq.~\eqref{redshift} in the main text, where we have dropped the zeroth-order contribution.

\section{\label{apdx3}Influence of Relative Motion Between Pulsar and Observer}

In the main text, we assumed the observer and pulsar were comoving for simplicity. 
In this appendix, we relax this approximation.
Relative motion between the pulsar and observer affects the observed timing residuals in two different ways: adding a contribution directly and modifying the time dependence of those produced by ultralight dark matter.
In this section, we examine these effects individually and show that it is a reasonable approximation to neglect the motion of the Earth around the Sun and regard the pulsar as comoving with the solar system.

The contributions to the redshift sourced by the relative motion between the Earth and the pulsar have been discussed in App.~\ref{apdx2}, where we have shown that they lead to the ordinary Doppler redshift as the zeroth-order term, $z^{(0)}$, and produce a correction of $\mathcal{O}(vh_{\mu\nu})$ to the first-order term, $z^{(1)}$, if $v$ is the typical velocity of the pulsar and the observer, with the superscript (0), (1) denoting respectively the unperturbed quantities and the first-order corrections in terms of $h_{\mu\nu}$.
However, the zeroth-order Doppler effect is a deterministic signal that does not contribute to the two-point correlation function, while the $\mathcal{O}(vh_{\mu\nu})$ correction to $z^{(1)}$ is negligible, since both the Earth and the pulsars are non-relativistic in the rest frame of the solar system where we calculate the correlation functions.

The relative motion alters the time dependence of timing residuals produced by ultralight dark matter and their two-point correlation functions due to the time-varying locations of the Earth and the pulsars.
The Earth rotates around the Sun at a speed $v_\mathrm{E} \simeq 10^{-4}$ within a range of $2\, \mathrm{AU} \simeq 9.8 \times 10^{-6} \mathrm{pc}$, much smaller than the Earth-pulsar distances.
The relative velocity of the pulsars to the Sun is typically given by the local velocity dispersion of stars, $v_\mathrm{P} \sim 10^{-3}$, which implies that the typical displacement of the pulsars with respect to the Sun over a detection time span $T \sim 10\, \mathrm{yr}$ is about $10^{-3}\, \mathrm{pc}$, also much smaller than their distances to the Earth.
Therefore, working in the rest frame of the solar system, we can safely neglect the variation of the pulsar orientations, $\hat{\mathbf{n}}$, and distances, $d$, over time in the two-point correlation functions, except for those in the imaginary exponents in $U_{ab}$ of the fast mode correlation function.

In these imaginary exponents, we also need to check the phase changes over time caused by the location variations. 
With the Earth and the pulsars in motion, the correlations of metric perturbations in the synchronous gauge that compose the fast mode timing residual two-point correlation function, $\left< r(t) r(t') \right>$, are generically:
\begin{align}
\begin{aligned}
    &\left< h_{ij}\big(t, \mathbf{x}_\mathrm{E}(t)\big) h_{kl}\big(t', \mathbf{x}_\mathrm{E}(t')\big) \right> \\
    -\,& \left<  h_{ij}\big(t, \mathbf{x}_\mathrm{E}(t)\big) h_{kl}\big(t'_{\mathrm{e}b}, \mathbf{x}_{\mathrm{P}b}(t'_{\mathrm{e}b})\big) \right> \\ 
    -\,& \left< h_{ij}\big(t_{\mathrm{e}a}, \mathbf{x}_{\mathrm{P}a}(t_{\mathrm{e}a})\big) h_{kl}\big(t', \mathbf{x}_\mathrm{E}(t')\big) \right> \\
    +\,& \left< h_{ij}\big(t_{\mathrm{e}a}, \mathbf{x}_{\mathrm{P}a}(t_{\mathrm{e}a})\big) h_{kl}\big(t'_{\mathrm{e}b}, \mathbf{x}_{\mathrm{P}b}(t'_{\mathrm{e}b})\big) \right>,
\end{aligned}
\end{align}
where $\mathbf{x}_\mathrm{E}(t)$ is the trajectory of the Earth, $\mathbf{x}_{\mathrm{P}a}(t)$, $\mathbf{x}_{\mathrm{P}b}(t)$ are the trajectories of pulsar $a$ and pulsar $b$, and $t_{\mathrm{e}a}$, $t'_{\mathrm{e}b}$ are the emission times of the photons from pulsar $a$ and pulsar $b$ that arrive at the Earth at respectively $t$ and $t'$, which solve the equations $t_{\mathrm{e}a} = t - |\mathbf{x}_{\mathrm{P}a}(t_{\mathrm{e}a}) - \mathbf{x}_\mathrm{E}(t)|$ and $t'_{\mathrm{e}b} = t' - |\mathbf{x}_{\mathrm{P}b}(t'_{\mathrm{e}b}) - \mathbf{x}_\mathrm{E}(t')|$.
Thus, $U_{ab}$ is modified to:
\begin{align}
\begin{aligned}
        U_{ab} &= e^{- |\mathbf{x}_\mathrm{E}(t) - \mathbf{x}_\mathrm{E}(t')|^2 / \ell^2}\, e^{i2m \left( - \bar{\mathbf{v}} \cdot \left(\mathbf{x}_\mathrm{E}(t) - \mathbf{x}_\mathrm{E}(t')\right) \right)} \\
        &- e^{- |\mathbf{x}_\mathrm{E}(t) - \mathbf{x}_{\mathrm{P}b}(t'_{\mathrm{e}b})|^2 / \ell^2}\, e^{i2m \left( |\mathbf{x}_{\mathrm{P}b}(t'_{\mathrm{e}b}) - \mathbf{x}_\mathrm{E}(t')| - \bar{\mathbf{v}} \cdot \left(\mathbf{x}_\mathrm{E}(t) - \mathbf{x}_{\mathrm{P}b}(t'_{\mathrm{e}b})\right) \right)} \\
        &- e^{- |\mathbf{x}_{\mathrm{P}a}(t_{\mathrm{e}a}) - \mathbf{x}_\mathrm{E}(t')|^2 / \ell^2}\, e^{i2m \left( - |\mathbf{x}_{\mathrm{P}a}(t_{\mathrm{e}a}) - \mathbf{x}_\mathrm{E}(t)| - \bar{\mathbf{v}} \cdot \left(\mathbf{x}_{\mathrm{P}a}(t_{\mathrm{e}a})- \mathbf{x}_\mathrm{E}(t')\right) \right)} \\
        &+ e^{- |\mathbf{x}_{\mathrm{P}a}(t_{\mathrm{e}a}) - \mathbf{x}_{\mathrm{P}b}(t'_{\mathrm{e}b})|^2 / \ell^2}\, e^{i2m \left( |\mathbf{x}_{\mathrm{P}b}(t'_{\mathrm{e}b}) - \mathbf{x}_\mathrm{E}(t')| - |\mathbf{x}_{\mathrm{P}a}(t_{\mathrm{e}a}) - \mathbf{x}_\mathrm{E}(t)| - \bar{\mathbf{v}} \cdot \left( \mathbf{x}_{\mathrm{P}a}(t_{\mathrm{e}a}) - \mathbf{x}_{\mathrm{P}b}(t'_{\mathrm{e}b}) \right) \right)},
    \end{aligned}
\end{align}
with additional time dependence contained in $\mathbf{x}_\mathrm{E}(t)$, $\mathbf{x}_{\mathrm{P}a}(t)$, and $\mathbf{x}_{\mathrm{P}b}(t)$ compared with Eq.~\eqref{U_ab}.
In the imaginary exponents, the dominant time dependence arises in the variation of propagation durations, $|\mathbf{x}_{\mathrm{P}b}(t'_{\mathrm{e}b}) - \mathbf{x}_\mathrm{E}(t')|$ and $|\mathbf{x}_{\mathrm{P}a}(t_{\mathrm{e}a}) - \mathbf{x}_\mathrm{E}(t)|$, which give rise to a typical phase change over the detection time span about $2mv_\mathrm{P}T \sim 10^{-2} (m/10^{-23} \mathrm{eV})$.
The phase changes that result from the variation of spatial arguments of the metric perturbations are suppressed by an additional $\bar{\mathbf{v}}$ and are about $2m|\bar{\mathbf{v}}|v_\mathrm{P}T \sim 10^{-5} (m/10^{-23} \mathrm{eV})$.
Both effects are negligible in the fast mode mass window.

In conclusion, the motion of the pulsars and the Earth relative to the Sun is completely irrelevant to our formalism and analysis, regardless of the precision of the pulsar location measurements.

\section{\label{apdx4}Eigenvalue Problem of the Fast Mode Timing Residual Covariance Matrix}

In this appendix, we prove that the covariance matrix defined by a monochromatic, stationary timing residual two-point correlation function on a discrete time series has at most $2N_\mathrm{P}$ nonzero eigenvalues if there are $N_\mathrm{P}$ pulsars involved in the analysis, and each eigenvector that corresponds to a nonzero eigenvalue is defined by a set of trigonometric functions.
The fast mode timing residual two-point correlation function Eq.~\eqref{fast mode timing residual correlation fucntion final result} and its defined covariance matrix conform to this case.
The proof naturally provides a simple method to compute the nonzero eigenvalues and their corresponding eigenvectors.

Consider a monochromatic timing residual correlation function in the form of Eq.~\eqref{fast mode timing residual correlation fucntion final result} with a generic angular frequency $\omega$.
In order to preserve the generality of our discussion, we do not specify the forms of $\mathcal{A}_{ab}$ and $U_{ab}$, only asking $\mathcal{A}_{ab}$ to be real and symmetric and $U_{ab}$ to be complex and Hermitian, i.e.,
\begin{align}
    \mathcal{A}_{ab} = \mathcal{A}_{ba}, \quad U^*_{ab} = U_{ba}.
\end{align}
Suppose for each pulsar $a$, $a=1,2,...,N_\mathrm{P}$, its timing residuals are measured on a series of arrival times $\{ t_{a,i} \}$, $i = 1,2,...,n_a$, which is not necessarily uniformly sampled.
The covariance matrix defined in this timing residual data vector space is:
\begin{align}\label{covariance matrix}
    C_{(a,i)(b,j)} \equiv \left< r_a(t_{a,i}) r_b(t_{b,j}) \right> = \mathcal{A}_{ab} \mathrm{Re} \Big[ U_{ab}\, e^{i\omega (t_{a,i} - t_{b,j})} \Big],
\end{align}
with $a,b = 1,2,...,N_\mathrm{P}$, $i=1,2,...,n_a$ and $j=1,2,...,n_b$.
The eigenvalue equation of this covariance matrix is therefore:
\begin{align}\label{eigenvalue problem}
\begin{aligned}
    (C\xi)_{a,i} \equiv \sum_{b=1}^{N_\mathrm{P}} \sum_{j=1}^{n_b} C_{(a,i)(b,j)} \xi_{b,j} = \sum_{b=1}^{N_\mathrm{P}} \sum_{j=1}^{n_b} \mathcal{A}_{ab} \mathrm{Re} \Big[ U_{ab}\, e^{i\omega (t_{a,i} - t_{b,j})} \Big] \xi_{b,j} = \lambda \xi_{a,i}\ ,
\end{aligned}
\end{align}
with $\lambda$ the eigenvalue and $\xi_{a,i}$ the eigenvector.
The last equality implies that if the eigenvalue $\lambda$ is nonzero, the eigenvector $\xi_{a,i}$ can definitely be written into the form of a trigonometric function of $t_{a,i}$ with the angular frequency $\omega$. 
We thus define:
\begin{align}\label{eigenvector as trigonometric functions}
    \xi_{a,i} = \mathrm{Re} \bigg[ \frac{\alpha_a}{\sqrt{n_a}} e^{i\omega t_{a,i}} \bigg], \quad \alpha_a \in \mathbb{C}, \quad a=1,2,...,N_\mathrm{P}.
\end{align}
It is worth noting that the same discrete vector $\xi_{a,i}$ can also be expressed by sampling other functions of time onto the time series $\{t_{a,i}\}$, but since time series are discrete and finite, these expressions are all equivalent.

Plugging this form of the eigenvector, Eq.~\eqref{eigenvector as trigonometric functions}, into the eigenvalue equation, Eq.~\eqref{eigenvalue problem}, we obtain:
\begin{align}
\begin{aligned}
    (C\xi)_{a,i} &= \sum_{b=1}^{N_\mathrm{P}} \sum_{j=1}^{n_b} \mathcal{A}_{ab} \mathrm{Re} \Big[ U_{ab}\, e^{i\omega (t_{a,i} - t_{b,j})} \Big] \mathrm{Re} \bigg[ \frac{\alpha_b}{\sqrt{n_b}} e^{i\omega t_{b,j}} \bigg] \\
    &= \frac{1}{2} \mathrm{Re} \bigg[ e^{i\omega t_{a,i}} \sum_{b=1}^{N_\mathrm{P}} \mathcal{A}_{ab} U_{ab} \sqrt{n_b} \bigg( \alpha_b + \frac{\alpha_b^*}{n_b} \sum_{j=1}^{n_b} e^{-i2\omega t_{b,j}} \bigg) \bigg] \\
    &= \lambda\, \mathrm{Re} \bigg[ \frac{\alpha_a}{\sqrt{n_a}} e^{i\omega t_{a,i}} \bigg]\ .
\end{aligned}
\end{align}
In the last equality which is the eigenvalue equation, as long as for each pulsar $a = 1,2,...,N_\mathrm{P}$, there are at least 2 measurements, i.e., $n_a \geq 2$, and the arrival times are not sampled such that the intervals between them are all integer multiples of $\pi/\omega$, we can select two arrival times, $t_{a,i_1}$ and $t_{a,i_2}$, that satisfy $e^{i2\omega (t_{a,i_1} - t_{a,i_2})} \neq 1$ and set up a system of equations, which yields:
\begin{align}\label{simplified eigenvalue problem}
    \left\{
    \begin{aligned}
        \frac{1}{2} \sum_{b=1}^{N_\mathrm{P}} \mathcal{A}_{ab} U_{ab} \sqrt{n_a n_b} \bigg( \alpha_b + \alpha_b^* \frac{1}{n_b} \sum_{j=1}^{n_b} e^{-i2\omega t_{b,j}} \bigg) = \lambda \alpha_a \\
        \frac{1}{2} \sum_{b=1}^{N_\mathrm{P}} \mathcal{A}_{ab} U^*_{ab} \sqrt{n_a n_b} \bigg( \alpha^*_b + \alpha_b \frac{1}{n_b} \sum_{j=1}^{n_b} e^{+i2\omega t_{b,j}} \bigg) = \lambda \alpha^*_a
    \end{aligned}
    \right. \ , \quad a = 1,2,...,N_\mathrm{P}.
\end{align}
The equations can be rewritten into a neater form by defining a vector $\vec{\alpha}$ with $\alpha_a$ the $a$-th element and defining two matrices $\mathbf{P}$ and $\mathbf{Q}$ with their elements given by
\begin{align}
    \mathbf{P}_{ab} &= \frac{1}{2} \mathcal{A}_{ab} U_{ab} \sqrt{n_a n_b}\ ,\\
    \mathbf{Q}_{ab} &= \frac{1}{2} \mathcal{A}_{ab} U_{ab} \frac{\sqrt{n_a}}{\sqrt{n_b}} \sum_{j=1}^{n_b} e^{-i2\omega t_{b,j}}\ .
\end{align}
These reorganize Eq.~\eqref{simplified eigenvalue problem} into a compact matrix form:
\begin{align}\label{matrix form eigenvalue problem}
    \left(
    \begin{matrix}
        \mathbf{P} & \mathbf{Q} \\
        \mathbf{Q}^* & \mathbf{P}^*
    \end{matrix}
    \right)
    \left(
    \begin{matrix}
        \vec{\alpha} \\
        \vec{\alpha}^*
    \end{matrix}
    \right) = \lambda
    \left(
    \begin{matrix}
        \vec{\alpha} \\
        \vec{\alpha}^*
    \end{matrix}
    \right)\ .
\end{align}
We recognize it as the eigenvalue equation of the $2N_\mathrm{P} \times 2N_\mathrm{P}$ matrix:
\begin{align}
    \mathbf{M} = 
    \left(
    \begin{matrix}
        \mathbf{P} & \mathbf{Q} \\
        \mathbf{Q}^* & \mathbf{P}^*
    \end{matrix}
    \right),
\end{align}
since it can be proved that all the eigenvectors of $\mathbf{M}$ can be chosen to be in the form of $(\vec{\alpha}^T, \vec{\alpha}^{\dagger})^T$.

Eq.~\eqref{matrix form eigenvalue problem} implies that, restricted to nonzero eigenvalues, the eigenvalue problem of the covariance matrix, Eq.~\eqref{covariance matrix}, is equivalent to the eigenvalue problem of $\mathbf{M}$ which has at most $2N_\mathrm{P}$ nonzero eigenvalues.
More specifically, supposing $\mathbf{M}$ has a total of $r \leq 2N_\mathrm{P}$ nonzero eigenvalues, $\lambda^1,\lambda^2,...,\lambda^r$, with their corresponding eigenvectors
\begin{align}
    \left(
    \begin{matrix}
        \vec{\alpha}^1 \\
        \vec{\alpha}^{1*}
    \end{matrix}
    \right), \
    \left(
    \begin{matrix}
        \vec{\alpha}^2 \\
        \vec{\alpha}^{2*}
    \end{matrix}
    \right),\ ...\ ,\
    \left(
    \begin{matrix}
        \vec{\alpha}^r \\
        \vec{\alpha}^{r*}
    \end{matrix}
    \right) \ , 
\end{align}
then all of the nonzero eigenvalues of the covariance matrix, Eq.~\eqref{covariance matrix}, are exactly $\lambda^1,\lambda^2,...,\lambda^r$ and their corresponding eigenvectors are
\begin{align}
    \xi^s_{a,i} = \mathrm{Re} \bigg[ \frac{\alpha^s_a}{\sqrt{n_a}} e^{i\omega t_{a,i}} \bigg], \quad s=1,2,...,r \; ,
\end{align}
with $\alpha^s_a$ the $a$-th element of $\vec{\alpha}^s$.
Thus, regardless of the size of the covariance matrix due to the large number of arrival times, it always has at most $2N_{\mathrm{P}}$ nonzero eigenvalues, which, along with their corresponding eigenvectors, can be extracted by simply solving the eigenvalue problem of the much smaller $2N_\mathrm{P} \times 2N_\mathrm{P}$ matrix $\mathbf{M}$.
The exact number of the nonzero eigenvalues depends on the specific forms of $\mathcal{A}_{ab}$ and $U_{ab}$ and the specific sampling of the arrival times, $\{t_{a,i}\}$.

In the limit that for every pulsar $b$, the observation time span is long enough, the number of measurements is large enough, and the sampling of the arrival times is uniform enough, the summation $\sum_{j=1}^{n_b} e^{-i2\omega t_{b,j}}$ vanishes such that $\mathbf{Q}$ vanishes.
The two-point correlation function that defines the covariance matrix as a functional operator in the continuous signal space also aligns with this case.
$\mathbf{M}$ thus becomes a Hermitian block diagonal matrix:
\begin{align}
    \mathbf{M} = 
    \left(
    \begin{matrix}
        \mathbf{P} & 0 \\
        0 & \mathbf{P}^*
    \end{matrix}
    \right),
\end{align}
and all of the eigenvalues of $\mathbf{M}$ must also be the eigenvalues of $\mathbf{P}$, which has at most $N_\mathrm{P}$ nonzero eigenvalues.
However, each eigenvector $\vec{\alpha}$ of $\mathbf{P}$, up to a complex coefficient, corresponds to two linearly independent eigenvectors of $\mathbf{M}$: 
\begin{align}
    \left(
    \begin{matrix}
        \vec{\alpha} \\
        \vec{\alpha}^{*}
    \end{matrix}
    \right),\ \text{and}\
    \left(
    \begin{matrix}
        i\vec{\alpha} \\
        -i\vec{\alpha}^{*}
    \end{matrix}
    \right) ,
\end{align}
and two linearly independent eigenvectors of the covariance matrix:
\begin{align}\label{1 to 2 map of eigenvectors}
    \xi_{a,i} = \mathrm{Re} \bigg[ \frac{\alpha_a}{\sqrt{N_a}} e^{i\omega t_{a,i}} \bigg]\ , \quad \text{and}\quad \xi'_{a,i} = \mathrm{Re} \bigg[ \frac{i\alpha_a}{\sqrt{N_a}} e^{i\omega t_{a,i}} \bigg]\ .
\end{align}
Therefore, in this case, each eigenvalue of $\mathbf{P}$ corresponds to 2 identical eigenvalues of $\mathbf{M}$ and the covariance matrix, which means the $r \leq 2N_\mathrm{P}$ eigenvalues of $\mathbf{M}$ and the covariance matrix are at least pairwise equal.
The computation of the corresponding eigenvectors of the covariance matrix can be further simplified by just solving for the eigenvectors of $\mathbf{P}$ and plugging them into Eq.~\eqref{1 to 2 map of eigenvectors}.

The conclusions in this appendix are general and are valid for any covariance matrix defined by a monochromatic, stationary correlation function and thus can be taken advantage of in the analysis of monochromatic timing residual signals from any kind of sources.

\end{document}